"D. GHITU" Institute of Electronic Engineering and Nanotechnologies
Stockholm University
The University of Twente

# THE 12TH INTERNATIONAL CONFERENCE ON INTRINSIC JOSEPHSON EFFECT AND HORIZONS OF SUPERCONDUCTING SPINTRONICS ABSTRACT BOOK

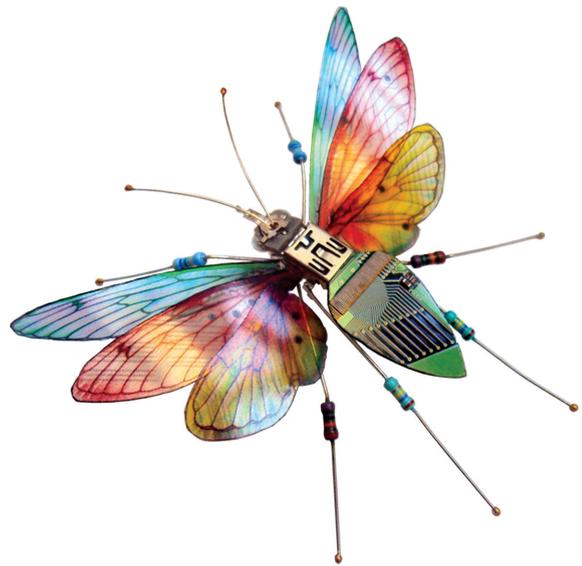

Chisinau, 2021

International Conference on Intrinsic Josephson Effect and Horizons of Superconducting Spintronics (12 ; 2021 ; Chişinău). The 12th International Conference on Intrinsic Josephson Effect and Horizons of Superconducting Spintronics, 22-25 September 2021, Chisinau, Moldova : Abstract Book / responsabil de ediţie: Anatolie Sidorenko ; organizing committee: Anatolie Sidorenko (Moldova), Alexander  Golubov (Netherlands), Vladimir Krasnov (Sweden). – Chişinău : S. n., 2021 (F.E.-P. "Tipografia Centrală"). – 87 p. : fig.

Antetit.: Inst. of Electronic Engineering and Nanotechnologies, Stockholm Univ., The Univ. of Twente. – Text parţial în lb. rom. – Referinţe bibliogr. la sfârşitul art. – Supported by the European Union. – 50 ex.

ISBN 978-9975-47-215-9.

538.9:620.3(082)

I-58

**Dear Participants of the IJE-21 Conference,**

The aim of the event is to strengthen regional and interdisciplinary links between the scientists, while supporting the development of international contacts and introducing young researchers to professional and fruitful scientific work practices, to bring together leading experts to share their expertise and experience in developing of new ideas and principles, novel technologies and their implementations.

**Main topics of the IJE-21 conference are:**

• Physics and applications of the intrinsic Josephson effect;
• S/F hybrid structures and horizons of superconducting spintronics;
• High-frequency Josephson devices;
• Modelling of functional nanostructures;
• Unconventional and topological superconductivity;
• Artificial Neural Networks, Qubits and quantum computing.

**Conference format**

Due to restrictions related to COVID-19 pandemics, the conference is organized in a mixed format both in physical and virtual space. Physically it takes place in Chisinau, Moldova - participants will get a regular service including coffee breaks and social activity (external session on 23 September, city tour and conference dinner). All presentations will be live-streamed, facilitating virtual participation.

Types of presentations: Oral (invited and regular, either physical or virtual), Poster (virtual only).

Best reports will be published as the articles in a special issue of the Beilstein Journal of Nanotechnology:

**Intrinsic Josephson effect and prospects of superconducting spintronics**
The editors: Anatoli Sidorenko, Vladimir Krasnov and Horst Hahn.

Details one can find on the page: https://www.beilstein-journals.org/bjnano/series/89

Wishing you a successful work at the IJE-21 and enjoyable stay in Moldova.

Director of the Conference

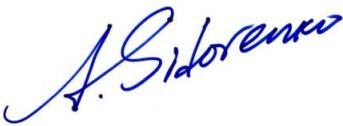

**Prof. Dr. Sidorenko Anatolie,**
Director of the conference, coordinator of the "SPINTECH" project,
Member of the Academy of Scientists of Moldova,
Chair of the Moldavian Physical Society (SFM),
Institute of Electronic Engineering and Nanotechnologies "D. GHITU"
Academiei Str. 3/3, MD2028, Chisinau, Moldova
Phone/Mob/Fax/E-mail: +373-22-737072/ +37369513294 /
+373-22-727088 / anatoli.sidorenko@kit.edu

The project "SPINTECH" has received funding from the European Union's Horizon 2020 research and innovation programme under grant agreement No 810144

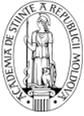 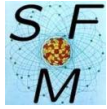 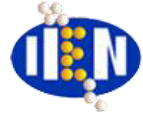 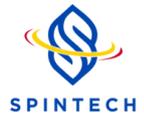

# THE 12TH INTERNATIONAL CONFERENCE ON INTRINSIC JOSEPHSON EFFECT AND HORIZONS OF SUPERCONDUCTING SPINTRONICS

## 22-25 September 2021, Chisinau, Moldova

This event is supported by the European Union H2020-WIDESPREAD-05-2017-Twinning project "SPINTECH" under grant agreement Nr. 810144

### ORGANIZING COMMITTEE

Prof. Anatolie Sidorenko, IEEN Chisinau, Moldova
Prof. Alexander Golubov, University of Twente, The Netherlands
Prof. Vladimir Krasnov, University of Stockholm, Sweden

### LOCAL ORGANIZING COMMITTEE

**Director of the conference - Prof. Anatolie Sidorenko**
"D. GHITU" Institute of Electronic Engineering and Nanotechnologies, Academiei str. 3/3, MD2028, Chisinau, Moldova, E-mail: anatoli.sidorenko@kit.edu, Phone: +37322737092/+37322727088
**Scientific secretary of the conference - Dr. Evgheni Antropov**
"D. GHITU" Institute of Electronic Engineering and Nanotechnologies, Academiei str. 3/3, MD2028, Chisinau, Moldova, E-mail: aidjek@gmail.com, Phone/Viber: +37379757615
**Administrator of the conference - Dr. Oleg Bujor**
"D. GHITU" Institute of Electronic Engineering and Nanotechnologies, Academiei str. 3/3, MD2028, Chisinau, Moldova, E-mail: bujor.oleg@gmail.com, Phone/Viber: +37379622073

# PROGRAM

## THE 12TH INTERNATIONAL CONFERENCE ON INTRINSIC JOSEPHSON EFFECT AND HORIZONS OF SUPERCONDUCTING SPINTRONICS

22-25 September 2021, Chisinau, Moldova

| DAY 1, Wednesday 22 September | | |
|---|---|---|
| 9:30-9:50 | Sidorenko Anatolie | Welcome note |
| **Session - I. Bi-2212 THz sources** (East Asia) | | |
| 9:50-10:20 | Kadowaki Kazuo | Towards Higher Power THz Emission from High -$T_c$ Superconducting $Bi_2Sr_2CaCu_2O_{8+d}$ IJJ Mesa Devices |
| 10:20-10:50 | Kakeya Itsuhiro | Mutual synchronization and helicity switching of terahertz radiation from intrinsic Josephson junction stack demonstrated by polarization analysis |
| 10:50-11:20 | Wei Zihan | Tailoring surface Josephson junction of Bi2Sr2CaCu2O8+δ for microwave applications. |
| 11:20-11:40 | Kuwano Genki | Mesa Sidewall Effects on Synchronization of Intrinsic Josephson Junctions in High-Tc Superconducting Bi2Sr2CaCu2O8+d Terahertz Emitters |
| 11:40-12:00 | Kashiwagi Takanari | Study of fabrication processes of high temperature superconducting terahertz emitters |
| 12:00-12:20 | Kobayashi Ryota | Analysis of Frequency Entrainment of Mutual Synchronous Terahertz Oscillation in Bi2212 Mesa Array |
| 12:40-14:00 | Lunch break | |
| **Session - II. Novel devices** | | |
| 14:00-14:30 | Hilgenkamp Hans | Challenges in energy-efficient computing |
| 14:30-14:50 | Kitano Haruhisa | Nonlinear Bifurcation in Microwave-Induced Phase Escapes of Current-biased Intrinsic Josephson Junctions |
| 14:50-15:20 | Delfanazari Kaveh | Quantum Terahertz Electronics: from superconducting coherent THz light sources to ultracompact waveguides and photonic integrated circuitry with quantum materials |
| 15:20-15:40 | Truccato Marco | X-ray nanopatterning on Bi-2212 whiskers |
| 15:40-16:00 | Krasnov Vladimir | Power efficient THz sources based on Bi-2212 whiskers |
| 16:00-16:30 | Klemm Richard | The important role of symmetry in the terahertz emission from thin Bi2Sr2CaCu2O8+δ microstrip antennas |
| 16:30-17:00 | Coffee break | |
| **Session - III. Layered superconductors** (USA) | | |
| 17:00-17:20 | Kim Philip | Toward realization of novel superconductivity based on twisted van der Waals Josephson junction in cuprates |
| 17:20-17:40 | Benseman Timothy | Intrinsic Josephson junction Bi2Sr2CaCu2O8 terahertz sources: Considerations for enhanced power output and operating temperature |
| 17:40-18:00 | Fomin Vladimir | Topological transitions in rolled-up superconductor nanomembranes under a strong transport current |
| 18:00-18:20 | Koshelev Alexei | Spin waves and related dynamic phenomena in layered superconductors with helical magnetic structure |
| 18:00-18:20 | Savva Yurii | Using the database and information technologies in studying functional nanostructures |
| 18:20-18:40 | Leonid Revin | Nonmonotonous temperature dependence of Shapiro steps in YBCO grain boundary junction |
| **DAY 2, Thursday 23 September** | | |
| **Session - IV. Unconventional superconductors and applications** | | |
| 9:30-10:00 | Kleiner Reinhold | Space-time crystalline order of a high-critical-temperature superconductor with intrinsic Josephson junctions |
| 10:00-10:30 | Shibauchi Takasada | Exotic superconducting states in FeSe-based materials |
| 10:30-11:00 | Lee Gil-Ho | Steady Floquet-Andreev States Probed by Tunnelling Spectroscopy |

| 11:00-11:20 | Li Qiang | Josephson coupling and pair density waves in stripe-ordered superconductors |
|---|---|---|
| 11:20-11:40 | Kieler Oliver | The pulse-driven AC Josephson voltage standard |
| 11:40-12:00 | Galin Mikhail | Fine Structure of Radiation Spectra from Large Josephson Systems |
| 12:00-12:30 | Kuzmin Leonid | Development of a Single Photon Counter for a Yoctojoule Energy Range |
| 12:30-14:00 | Lunch break | |
| 14:00-16:30 | **Poster session** | Cattaneo Roger, Chiginev Alexander, Krasnov Mihail, Kulikov Kirill, Novikova Nataliya, Pimanov Dmitry, Boian Vladimir, Wolter Silke, Kapran Olena, Mazanik Andrey, Muntyanu Fiodor, Rahmonov Ilhom, Severyukhina Olesya, Yanilkin Igor, Yusupov Roman, Bobkov Grigorii, Condrea Elena, Gutsul Tatiana, Morari Vadim, Suvorov Stepan, Sirbu Andrei, Delfanazari Kaveh, Chornopyshchuk Roman, Kim Philip, Konopko Leonid, Lupu Maria, Nica Iurie N., Nikolaeva Albina, Pankratov Andrey L., Shibauchi Takasada, E. G. Coscodan, E. A. Zasavitsky |
| **External Session - V. (in Cricova)** | | |
| 14:00-18:00 | Round table discussion in Cricova | |
| **DAY 3, Friday 24 September** | | |
| **Session - VI. Unconventional superconductors** | | |
| 9:30-10:00 | Kordyuk Alexander | Multiband electronic structure as a key to high temperature superconductivity and new quantum applications |
| 10:00-10:30 | Belzig Wolfgang | Magnetic Field-Induced "Mirage" Gaps and Triplet Josephson Effect in Ising Superconductors |
| 10:30-10:50 | Vakhrushev Alexander | Modeling of technological processes for the formation of nanostructures on a solid surface |
| 10:50-11:10 | Shukrinov Yury | Phase dynamics, IV-characteristics and magnetization dynamics of the φ0 Josephson junction |
| 11:10-11:30 | Kusmartsev Fiodor | Topological Order in HighTemperature Superconductors |
| 11:30-11:50 | Pankratov Andrei | Development of Dichroic 220/240 GHz Parallel Arrays of Cold-Electron Bolometers with Slot Antennas for LSPE project |
| 11:50-12:10 | Karabassov Tairzhan | Reentrant superconductivity in proximity to a topological insulator |
| 12:10-14:00 | Lunch break | |
| **Session -VII. S/F** | | |
| 14:00-14:30 | Ryazanov Valery | Superconducting phase inversions in mesoscopic superconductor-normal metal-ferromagnet Josephson structures controlled by nonequilibrium |
| 14:30-15:00 | Klenov Nikolai | Superconducting neuron for networks based on radial basis functions |
| 15:00-15:30 | Tagirov Lenar/ Yanilkin Igor | Controllable two- vs three-state magnetization switching in single-layer epitaxial Pd1-xFex films and Pd0.92Fe0.08/Ag/Pd0.94Fe0.04 heterostructure |
| 15:30-15:50 | Sidorenko Anatolie | Nanostructures Superconductor/Ferromagnet for Superconducting Spintronics |
| 15:50-16:10 | Soloviev Igor | Approaches to scaling superconducting digital circuits |
| 16:10-16:30 | Bobkova Irina | Triplet superconductivity induced by moving condensate |
| 16:30-16:50 | Khaydukov Yuri | Chirality of Bloch domain walls in exchange biased CoO/Co bilayer seen by waveguide-enhanced neutron spin-flip scattering |
| 16:50-18:00 | Coffee break and closing remarks of the conference | |
| 19:30 | Conference dinner | |

# TABLE OF CONTENT

# Towards Higher Power THz Emission from High -Tc Superconducting Bi2Sr2CaCu2O8+d IJJ Mesa Devices

_K. Kadowaki[1]_, _Y. Ono[2]_, _H. Minami[1,2]_, _S. Kusunose[2]_, _T. Yuhara[2]_, _G. Kuwano[2]_, _S. Nakagawa[2]_, _J. Kim[2]_, _M. Nakayama[2]_, _K. Nagayama[2]_, _T. Kashiwagi[1,2]_, _M. Tsujimoto[1,3]_ and _R. A. Klemm[4]_

[1]_Faculty of Pure & Applied Sciences, University of Tsukuba, 1-1-1, Ten-nodai, Tsukuba, 305-8573, Japan_

[2]_Graduate School of Pure & Applied Sciences, University of Tsukuba, 1-1-1, Ten-nodai, Tsukuba, Ibaraki 305-8573, Japan_

[3]_Research Center for Emerging Computing Technologies, National Institute of Advanced Industrial Sciences and Technology (AIST), Central 2, 1-1-1, Umezono, Tsukuba, Ibaraki 305-8568, Japan_

[4]_Department of Physics, University of Central Florida, 4111 Libra Drive, Orlando, FL32816-2385 USA_

Electromagnetic waves (EMWs) with terahertz (THz) frequencies are a superseding target not only for the next generation communication such as 6G or 7G, but also emerging novel quantum computing and the related technologies. In particular, the development of continuous coherent THz sources with sufficient power and sensitive detectors are urgent technological issues in extending present available semiconductor as well as far-infrared optical device researches. In contrast, in 2007, we discovered the third novel method which can generate tunable, continuous, and coherent THz EMWs using high-$T_c$ superconducting $Bi_2Sr_2CaCu_2O_{8+d}$ intrinsic Josephson junctions (IJJs)[1]. Although the emission covers the most important THz region from about 0.3 to 2.4 THz with the line width of ∼0.1 GHz, it is rather difficult to have larger power more than ∼1 mW. So far, our effort has been made towards two directions: one is to establish emitters with higher efficiency, and the other is to enhance the power by external (internal) power amplifier. Devices having a separating THz source form the cavity resonator, or devices having impedance matching through various antenna structures, *etc*. have been tried along the former line, the results are yet far from the drastic improvement[2]. We are, therefore, shifting to the second strategic line taking external power amplifiers such as parametric as well as plasmonic origins are under consideration.

**Dr. Kazuo Kadowaki**
Faculty of Pure & Applied Sciences, University of Tsukuba
1-1-1, Ten-nodai, Tsukuba, 305-8573, Japan
E-mail: dr.kazuo.kadowaki@gmail.com

# Mutual synchronization and helicity switching of terahertz radiation from intrinsic Josephson junction stack demonstrated by polarization analysis

*I. Kakeya[1], K. Hayama[1], R. Kobayashi[1], K. Maeda[1], A. Elarabi[1\*], S. Fujita[1], G. Kuwano[2], M. Tsujimoto[2,3], and H. Asai[3]*

[1]*Department of Electronics Science and Engineering, Kyoto University, Kyoto, Japan*
[2]*Graduate School of Pure and Applied Sciences, University of Tsukuba, Tsukuba, Japan*
[3]*National Institute of Advanced Industrial Science and Technology (AIST), Tsukuba, Japan*
*\*Present address: Okinawa Institute of Science and Technology Graduate University, Onna, Japan*

In this talk, we discuss availability to describe the synchronization phenomena between two IJJ mesas with the bases of the two mesas individually biased. We estimated Stokes polarization parameters (SPPs) of individually and simultaneously biased mesas. SPPs are obtained from transmission intensity through a fixed wire grid polarizer and a quarter-wave plate rotating around the optic axis (polarization analyzer). The polarization ellipse for the synchronized emission can be reproduced by a sum of those of individual emissions with proper complex coefficients. This consideration can be described as $|1\&2\rangle = \alpha|1\rangle + \beta|2\rangle$, where $|1\&2\rangle$, $|1\rangle$ and $|2\rangle$ represent emitted electromagnetic waves from simultaneously and individually biased mesas such as $|1\rangle = E_1 e^{i(\omega_1 t + \delta_1)}$, where $\omega_1$ and $\delta_1$ represent frequency and phase of the plasma oscillation of the mesa 1, for example[1]. Figure 1 shows polarization ellipses of single mesas C and E, and the pair them parallelly connected, which are represented as $|C\rangle$, $|E\rangle$ and $|C \parallel E\rangle$, respectively. Here, the data are taken at the intensity maximum of $|C \parallel E\rangle$ and the local intensity maxima in the vicinity of the equivalent voltage of $|C\rangle$ and $|E\rangle$. This result arises $|\alpha| = 1.27$, $|\beta| = 0.54$, and $\arg(\beta/\alpha) = 56$ deg. Since the phase difference between $|C\rangle$ and $|E\rangle$ can be given by the separation between the mesas d, the equivalent phase different can be estimated as $\delta_d = 2\pi n d f_e / c_0$, where $f_e$ =0.5 THz and $n$=4.1 are the emission frequency and reflective index of the mesa, respectively. As a result, $\delta_d$ is estimated as 57 deg., which is very close to the $\arg(\beta/\alpha)$.

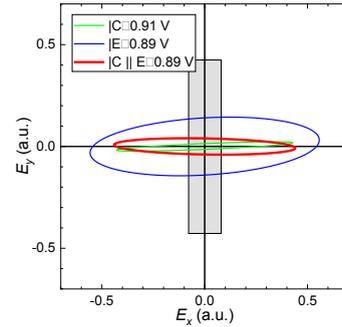

**Fig. 1.** Polarization ellipses for individually biased mesas C and E, and parallelly biased C and E at representative voltages

Circularly polarized electromagnetic waves are radiated from specifically designed mesa-devices [2]. Helicity of emitted electric field is determined by the chirality of the device. We have demonstrated to switch helicity by switching injection electrode.

**Prof. Dr. Itsuhiro Kakeya**
Kyotodaigaku-Katsura
615-8510 Kyoto, Japan
Phone: +81 75-383-2265
E-mail: kakeya@kuee.kyoto-u.ac.jp
Web: http://sk.kuee.kyoto-u.ac.jp/ja/

**Zihan Wei**

# Tailoring surface Josephson junction of Bi2Sr2CaCu2O8+δ for microwave applications

**Zihan Wei [1], Hongmei Du [1], Dingding Li [1], Shiyu Tong [1], Ping Zhang [1], Tianyuan Chi [1], Yangyang Lv [1], Hancong Sun [2], Huabing Wang [1,2*], Peiheng Wu [1,2]**

[1] Research Institute of Superconductor Electronics, Nanjing University, Nanjing 210023, China

[2] Purple Mountain Laboratories, Nanjing 211111, China

*E-mail: hbwang@nju.edu.cn

Superconducting properties of the surface intrinsic Josephson junctions of Bi2Sr2CaCu2O8+δ single crystal are pretty different from the inner ones. They can be used to detect microwave response at a relatively low frequency. In this paper, we systematically investigated and then tailored the surface Josephson junctions. The key technique is in-situ gold evaporation at room temperature, with which we can obtain a surface junction with desirable properties. The highest superconducting transition temperature ($T_c$) is about 89 K, even as high as that of the inner junctions. It shows a clear microwave response at low frequency of 20 GHz to 40 GHz from 3.1 K to 87 K. The lowest $T_c$ of a surface junction is about 32 K. Since it has a very small critical current, it works like a single junction over a broad current range without inner ones appearing. With these surface junctions, we have carried out various experiments on microwave response. Detailed experimental results will be reported at the conference.

## Acknowledgements

We gratefully acknowledge financial support by the National Natural Science Foundation of China (Grant Nos. 61727805, 11961141002, 61521001), and Jiangsu Key Laboratory of Advanced Techniques for Manipulating Electromagnetic Waves.

**Dr. Zihan Wei**
Research Institute of Superconductor Electronics
Nanjing University, Nanjing 210023, China
E-mail: hbwang@nju.edu.cn

# Mesa Sidewall Effects on Synchronization of Intrinsic Josephson Junctions in High-$T_c$ Superconducting Bi2Sr2CaCu2O8+δ Terahertz Emitters

_G. Kuwano_ [1], K. Nagayama [1], S. Shouhei [1], T. Yuhara [1], T. Kashiwagi [1,2], H. Minami [1,2], K. Kadowaki [2], M. Tsujimoto [2,3]

[1] Graduate School of Pure and Applied Sciences, University of Tsukuba, 1-1-1 Tennodai, Tsukuba, Ibaraki 305-8573, Japan

[2] Facutly of Pure and Applied Sciences University of Tsukuba, 1-1-1 Tennodai, Tsukuba, Ibaraki 305-8573, Japan

[3] Research Center for Emerging Computing Technologies, National Institute of Advanced Industrial Science and Technology (AIST), Central2, 1-1-1 Umezono, Tsukuba, Ibaraki 305-8568, Japan

A stack of intrinsic Josephson junctions (IJJs) of cuprate high-critical-temperature (high-$T_c$) superconductors has been shown to generate intense and coherent terahertz (THz = $10^{12}$ Hz) electromagnetic radiation [1,2]. So far, the intense radiation from mesa structures of IJJs was demonstrated using single crystalline Bi2Sr2CaCu2O8+δ, where its superconducting energy gap of a few tens of meV corresponds to approximately 12 THz. The intense emission was observed at the characteristic frequencies that fulfill the geometrical cavity resonance conditions [1,3]. Meanwhile, the most intriguing physics of this peculiar phenomenon is the mutual synchronization among thousands of stacked IJJs with distributed widths due to the trapezoidal cross-section profile: An IJJ mesa has a trapezoidal cross-section profile due to the limited accuracy of microfabrication processes, which may produce different emission characteristics.

In this study, we have established a method of accurate control of Bi-2212 mesa sidewall angle to investigate the effect of that on mutual synchronization among stacked IJJs [4]. We succeeded in determining a characteristic threshold for realization of coherent terahertz radiation. In this talk, we will present the details of the fabrication process and experimental results.

**Prof. Genki Kuwano**
Graduate School of Pure and Applied Sciences,
University of Tsukuba,
1-1-1   Tennodai, Tsukuba,
Ibaraki 305-8573, Japan
E-mail: d31018gk@gmail.com

## Study of fabrication processes of high temperature superconducting terahertz emitters

_T. Kashiwagi_ [1, 2], T. Imai [2], S. Nakagawa [2], M. Nakayama [2], J. Kim [2], T. Yamaguchi [2], G. Kuwano [2], K. Nagayama [2], T. Yuhara [2], Y. Saito [2], S. Suzuki [2], M. Tsujimoto [1,3], H. Minami [1,2], K. Kadowaki [1]

[1] Faculty of Pure & Applied Sciences, University of Tsukuba, Ibaraki, Japan

[2] Graduate School of Pure & Applied Sciences, University of Tsukuba, Ibaraki, Japan

[3] Research Center for Emerging Computing Technologies, AIST, Ibaraki, Japan

Single crystals of high-Tc superconductor $Bi_2Sr_2CaCu_2O_{8+\delta}$ (Bi2212) contain insulating Bi2O2 layers and superconducting $CuO_2$ layers. These layers are alternately stacked along the crystallographic c-axis and are known as intrinsic Josephson junctions (IJJs) [1, 2]. By processing Bi2212 single crystals into mesa structures and applying dc bias voltages across the IJJs, terahertz (THz) waves can be produced [3]. We developed Bi2212-THz emitters based on the stand-alone type of mesa structures (SAMs). A sandwich structure of SAM can remove self- generated heat from a mesa structure more efficiently and increase emission power and the radiation frequency range widely [4]. To obtain farther development of IJJ-devices, we have studied fabrication processes of SAMs using a wet etching method [5]. Thicker SAMs, especially, would help to obtain higher emission power according to the emission principle of this device. We investigated etching solutions for thicker SAMs up to about 10 µm. In addition, we also developed a fabrication method to obtain SAM chips for mass production of Bi2212- THz emitters. The method enables us to produce many thicker SAMs at a time and would be useful to improve emission power.

**Prof. Takanari Kashiwagi**
Faculty of Pure & Applied Sciences
University of Tsukuba
1-1-1 Tsukuba, Ibaraki, Japan
E-mail: kashiwagi@ims.tsukuba.ac.jp

Ryota Kobayashi

# Analysis of Frequency Entrainment of Mutual Synchronous Terahertz Oscillation in Bi2212 Mesa Array

R. Kobayashi[1], K. Hayama[1], S. Fujita[1], I. Kakeya[1]

[1] Kyoto University, Department of Electronic Science and Engineering, Kyoto, Japan

The high-temperature superconductor Bi2 Sr2CaCu2O8+δ (Bi2212) has a stacked structure of superconducting and insulating atomic layers inside the crystal, which is referred as an intrinsic Josephson junction (IJJ), and radiates electromagnetic waves in the terahertz range by the AC Josephson effect and cavity resonance effect when a voltage is applied to the mesa structure formed on the surface of a single crystal [1,2]. It has been reported that more powerful emission is obtained by operating multiple mesas formed on the same crystal substrate simultaneously [3]. The direct evidence that Josephson oscillations in two mesas are mutually synchronized through the superconducting substrate has been obtained from polarization measurements of oscillating electromagnetic waves in parallel bias of two mesas [4]. Understanding the inter-mesa coupling mechanism of this phenomenon is extremely important not only for understanding of many-body synchronization phenomenon of superconducting pendula but also for the realization of practical high-power terahertz light sources. It is difficult to reproduce the polarization characteristics during synchronous oscillation as a superposition of experimentally observed individual oscillation states of a mesa. This is because, in general, two oscillators in synchronous oscillation are affected by frequency retraction due to their interaction, and the oscillation state of each mesa deviates from that of the individual operation. In this analysis, the voltage variable polynomial approximation up to the second order of the Stokes parameter is used to extend the polarization characteristics of the single mesa oscillation in the terahertz oscillator with the mesa array structure to the vicinity of the measurement bias point. The result is shown in Figure 1. The extension of the discrete point to a continuous quantity makes it possible to reproduce the polarization characteristics during synchronous oscillation more precisely as a superposition of single oscillation states. Using the result, we discuss the interaction matrix elements between mesas that causes mutual synchronization in the Josephson oscillator, focusing on frequency modification and phase synchronization.

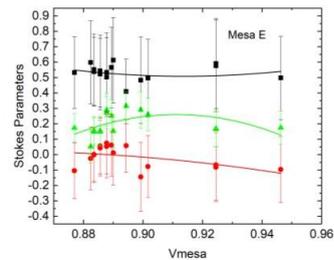

Fig. 1. Second order polynomial approximation of Stokes parameters by voltage variable

**Mr. Ryota Kobayashi**
Kyoto daigaku-katsura, Nishikyo-ku, Kyoto, Japan
Phonw : +81-75-383-2271
Fax: +81-75-383-2270
E-mail: kobayashi.ryota.86a@st.kyoto-u.ac.jp
Web: http://sk.kuee.kyoto-u.ac.jp/ja/

# Challenges in energy-efficient computing

**Prof. Dr. Ir. Hans Hilgenkamp**
University of Twente
Faculty of Science and Technology
Carré (building no. 15), room C2063
Hallenweg 23
7522NH Enschede
The Netherlands
E-mail: j.w.m.hilgenkamp@utwente.nl ; h.hilgenkamp@utwente.nl

# Nonlinear Bifurcation in Microwave-Induced Phase Escapes of Current-biased Intrinsic Josephson Junctions

**H. Kitano**, A. Yamaguchi, H. Ohnuma, Y. Watabe, S. Umegai, K. Hosaka

*Department of Physics, Aoyama Gakuin University, Sagamihara, Japan*

A better understanding of the complicated phase dynamics in the current-biased intrinsic Josephson junctions (IJJs) is crucially required for the implementation of IJJ-based THz emitter and superconducting qubit. Especially, the phase switches from the finite voltage state in IJJs are very interesting, since ac Josephson current generated in the phase-switched junction can induce a rapid oscillation of the potential describing the phase dynamics in another junction of IJJs [1], and the microwave-induced resonant phase escapes were observed up to a relatively high temperature [2]. Here, we report on the detailed analyses of the oscillation frequency $\omega_p$ in the potential well and the barrier height $\Delta U$ for the phase escapes as a function of the microwave power $P_{MW}$ irradiated to $Bi_2Sr_2Ca_{1-x}Y_xCu_2O_y$ IJJs [3]. We find that the microwave irradiation with a driving frequency lower than $\omega_p$ more largely decreases $\omega_p$ rather than $\Delta U$, showing a contrast to the theoretical model of quantum escapes in strongly driven Josephson junctions [4]. We argue that the microwave-induced decrease of $\omega_p$ is successfully explained by the harmonics of nonlinear bifurcation, discussed in terms of Josephson bifurcation amplifier [5]. A close similarity between the microwave irradiation and the generation of ac Josephson current in the voltage state strongly suggests that the nonlinear bifurcation with a large oscillation amplitude dominates the phase escapes from the finite voltage state of IJJs, occurring in the low temperature region.

**Prof. Dr. Haruhisa Kitano**
Fuchinobe 5-10-1, Chuo-ku
252-5258 Sagamihara, Kanagawa, Japan
Phone: +81 42-759-6286
Fax: +81 42-759-6444
E-mail: hkitano@phys.aoyama.ac.jp
Web: http://www.phys.aoyama.ac.jp/~w3-kitano/index_eg.html

**Kaveh Delfanazari**

## Quantum Terahertz Electronics: from superconducting coherent THz light sources to ultracompact waveguides and photonic integrated circuitry with quantum materials

*Kaveh Delfanazari[1,2]*

[1] *James Watt School of Engineering, University of Glasgow, UK*

[2] *Electrical Engineering Division & Cavendish Lab. University of Cambridge, UK*

Current compact emitter and receiver technologies are generally inefficient and impractical at terahertz (THz) frequencies between 0.1 and 10 THz. Hence, a gap exists between mature microwave and developed optical technologies. On-chip, integrated broadly tunable and powerful quantum sources that coherently radiate THz waves between 0.1 and 11 THz (potentially extendable to 15 THz) can be achieved based on quantum tunneling of electron pairs across the stack of intrinsic Josephson junctions (IJJs) naturally present in a single crystal of the layered high-temperature superconducting BSCCO. Such superconducting devices have been found to be especially promising solid-state THz sources capable of bridging the entire THz gap, as their wide-frequency tunability range is superior to that obtained from their semiconducting-based rivals, either single resonant-tunneling diodes (RTDs) or THz-quantum cascade lasers (QCLs).

I will review our recent results on superconducting THz devices and discuss our approaches towards their integration, scale up, and performance development. I will conclude by demonstrating novel low-loss photonic integrated circuits, waveguides, tunable modulators, slow light devices and filters.

**Dr. Kaveh Delfanazari**

Electronics and Nanoscale Engineering (ENE) Division, School of Engineering, University of Glasgow, Glasgow, G12 8QQ UK

E-mail: Kaveh.Delfanazari@glasgow.ac.uk

# X-ray nanopatterning on Bi-2212 whiskers

*M. Truccato[1,2]*

1. Department of Physics, University of Torino, via Giuria 1, 10125, Torino, Italy
2. I.N.F.N., Sezione di Torino, via Giuria 1, 10125 Torino Italy

Radiation damage is a well-known phenomenon that is of paramount importance at XFELs. However, the recent development of nanofocused hard X-ray beams at 3rd and 4th generation synchrotron radiation facilities has pushed the values of the instantaneous power density to limits where this problem must be considered in these experiments, too. A positive approach to this situation consists in exploiting the defects induced by X-ray irradiation in order to tailor the material properties in a desired way at the nanometric level. This direct-write X-ray nanopatterning (XNP) technique implies no etching and has already proved to be suitable to locally modify the electrical properties of both semiconducting and superconducting oxides. [1,2]

The typical experimental setup consists of a 17.5 keV beam, about $50 \times 50$ nm$^2$ in size and with a time-averaged photon flux of the order of $10^{11}$ photons per second, and in principle also allows for on-line electrical monitoring during device fabrication.[3] X-ray nanodiffraction experiments on $Bi_2Sr_2CaCu_2O_{8+\delta}$ (Bi-2212) and $YBa_2Cu_3O_{7-\delta}$ (Y-123) whisker-like single crystals have shown that these irradiations induce the appearance of grain boundaries with a corresponding increase of the crystal mosaicity, along with a decrease of the oxygen non-stoichiometric content $\delta$, which shifts the material doping status towards the underdoped regime. It has also been proved that an exponential relationship exists between the material resistivity and the irradiation dose or fluence, and that relaxation of pre-existing stress can take place during irradiation. Presently, the microscopic mechanisms implied in XNP have been not elucidated yet. According to numerical simulations, the heating induced by the nanobeam cannot be responsible for ordinary melting, and also photogenerated electrons knocking on the loosely-bound Bi-2212 interstitial oxygen atoms can only play a minor role. Nevertheless, XNP has already been used to fabricate stacks of intrinsic Josephson junctions out of Bi-2212 whiskers [4]. In principle, this technique also allows the fabrication of many in-series stacks of Josephson junctions, which could increase the chances for the observation of superradiance.

**Marco Truccato**
University of Torino, via Giuria 1, I-10125
Torino, Italy
E-mail: marco.truccato@unito.it

## Power efficient THz sources based on Bi-2212 whiskers

*R. Cattaneo[1], E. A. Borodianskyi[1], A. A. Kalenyuk[1,2], K. I. Shiianov[3,1], A. E. Efimov[3,1], N.D. Novikova[4,3], M. M. Krasnov[4,3], A. Agostino[5], M. Truccato[5] and V. M. Krasnov[1,3]*

[1] *Stockholm University,Physics Department, SE-10691 Stockholm, Sweden*

[2] *Institute of Metal Physics of National Academy of Sciences of Ukraine, 03142 Kyiv, Ukraine*

[3] *Moscow Institute of Physics and Technology, 141700 Dolgoprudny, Russia*

[4] *Keldysh Institute of Applied Mathematics, Miusskaya sq. 4, 125047 Moscow, Russia*

[5] *Department of Physics, NIS Interdepartmental Centre, University of Torino and Italian Institute of Nuclear Physics, 10125 Torino, Italy*

Low power efficiency is one of the main problems of THz sources, colloquially known as "the THz gap". Although a big progress was achieved in development of semiconducting quantum cascade lasers, their efficiency drops from 28% at 55 THz to a sub-percent at 3-4 THz and to ~0.01% at 1.3 THz.

Here we present prototypes of novel THz devices based on whisker-crystals of a high-temperature superconductor $Bi_2Sr_2CaCu_2O_{8+\delta}$. We employ various on-chip and off-chip detection techniques and, in particular, use the radiative cooling phenomenon for accurate evaluation of the emission power. It reveals that the efficiency of our devices can reach 12%. This is more than an order of magnitude larger than for similar devices made on conventional large-size crystals and not far away from the theoretical limit of 50% [1]. The boost of efficiency is attributed to a specific turnstile-antenna-like geometry of devices, which allows a drastic reduction of the parasitic crystal/electrode capacitance and enables good impedance matching with free space. We argue that such devices can be used for creation of tunable, monochromatic, continuous-wave, compact and power-efficient THz sources.

### Acknowledgements

The work was supported by the Russian Science Foundation Grants 19-19-00594 (experiment) and 20-42-04415 (sample fabrication).

**Prof. Vladimir Krasnov**
Stockholm University,Physics Department, SE-10691 Stockholm, Sweden
E-mail: vladimir.krasnov@fysik.su.se
Phone: +46-8-55378606

Richard A. Klemm

# The important role of symmetry in the terahertz emission from thin Bi2Sr2CaCu2O8+δ microstrip antennas

*Richard A. Klemm[1,2], Ruqayyah Shouk[1], Joseph R. Rain[1], PeiYu Cai[1,3], Alexander Baekey[1], Matthew A. Reinhard[1,4], Roman I. Vasquez[1,5], Andrew C. Silverman[1], and Christopher L. Cain[1]*

[1] Department of Physics, University of Central Florida, 4111 Libra Drive, Orlando, Florida 32816-2385 USA

[2] U. S. Research Laboratory, Wright-Patterson Air Force Base, Ohio 45433-7251, USA

[3] Imperial College London, Department of Physics, South Kensington Campus, London SW7 2AZ, UK

[4] Department of Physics, University of Florida, 2001 Museum Road, P. O. Box 118440, Gainesville, Florida 32611-8440, USA

[5] Department of Mathematics and Statistics, Auburn University, 221 Roosevelt Concourse, Auburn, Alabama 36849, USA

For thin microstrip antennas of Bi2Sr2CaCu2O8+δ, the spatial symmetry of the antenna is an important feature that has often been overlooked in the literature. In the high-symmetry cases of a disk, square, or equilateral triangle, the wave functions with the appropriate Neumann boundary conditions are grouped into two distinct classes, which are one- and two-dimensional representations of the respective point groups $C_{\infty v}$, $C_{4v}$, and $C_{3v}$. We present visualizations of representative wave functions of both of these distinct classes for each of these point groups. For the one-dimensional representations, color contour plots are presented. For the two-dimensional representations, the symmetry operations are worked out for the most general forms, and the infinite degeneracies are presented as sets of common nodal points, the patterns of which are invariant under all operations of the respective point group. In each of these high-symmetry shapes, the lowest energy wave function is a two-dimensional representation, which is infinitely degenerate. This implies that an ideal thermally managed, stand- alone mesa with one of these high symmetries will not exhibit a cavity resonance at its fundamental frequency. Therefore, in order to design arrays of Bi2Sr2CaCu2O8+δ microstrip antennas for high output power, it is essential for the array components to have a significant number of non-degenerate, low- energy wave functions. Two possible array designs will be presented and discussed.

**Prof. Richard A. Klemm**
Department of Physics,
University of Central Florida,
4111 Libra Drive, Orlando,
Florida 32816-2385 USA

## Toward realization of novel superconductivity based on twisted van der Waals Josephson junction in cuprates

*Philip Kim[1]*
*[1]Department of Physics, Harvard University, Cambridge MA 02138, USA*

Engineering moire superlattices by twisting and stacking two layers of Van der Waals materials has proved to be an effective way to promote interaction effects and induce exotic phases of matter. The high-temperature superconductor $Bi_2Sr_2CaCu_2O_{8+x}$ (Bi-2212) represents a prototypical cuprate superconductor, with weakly bonded van der Waals (vdW) layers. Using a novel cryogenic van der Waals pickup technique, we have fabricated Josephson junctions between two exfoliated Bi-2212 crystals with controlled relative twist angles. To preserve the air- and heat-sensitive Bi-2212 surface's integrity, we handled the devices entirely within argon or high-vacuum environment and kept the devices cold throughout the fabrication process. The resulting junctions support a Josephson critical current density of similar magnitude as the bulk c-axis intrinsic junctions and with Tc within a few Kelvin of the bulk value. With no need of a post-stacking anneal step, the interface shows minimal signs of surface degradation or reconstruction. The junctions' critical current evolves as expected for a d-wave superconductor. Our new fabrication methods open the possibility of creating arbitrarily complex, monolayer Bi-2212 heterostructures. I will discuss the most recent experimental results that hint novel superconducting states appeared at the twisted interface of cuprates vdW Josephson junctions.

**Prof. Philip Kim**
11 Oxford Street, Harvard University, Cambridge MA 02138, USA
Phone: +1-617-496-0714
E-mail: pkim@physics.harvard.edu
Web: https://kim.physics.harvard.edu//

# Intrinsic Josephson junction $Bi_2Sr_2CaCu_2O_8$ terahertz sources: Considerations for enhanced power output and operating temperature

*K. J. Kihlstrom[1], K. C. Reddy[1], A. E. Koshelev[2], U. Welp[2], W.-K. Kwok[2], K. Kadowaki[3], T. M. Benseman[1,2]*

[1]*CUNY Queens College, Department of Physics, New York City, New York, U.S.A.*
[2]*Argonne National Laboratory, Materials Science Division, Argonne, Illinois, U.S.A*
[3]*University of Tsukuba, Institute for Materials Science, Tsukuba, Japan*

The high-temperature superconductor $Bi_2Sr_2CaCu_2O_8$ contains stacked 'intrinsic' Josephson junctions, with very high packing density and a large superconducting gap energy. Rectangular 'mesa' devices constructed from this material are consequently a promising technology for coherent, continuous-wave radiation in the 'terahertz gap' range, spanning from approximately $0.3 - 2.0$ THz. A key issue for technological applications of such devices is their cryocooling requirements, and it is therefore highly desirable to optimize their performance at temperatures that can be achieved while using liquid nitrogen cryogenics, or by using highly compact Stirling micro-cryocoolers. Here we report 0.13 milliwatts of coherent emission power at 0.5 THz, at a bath temperature of 77 Kelvin. We achieved this by exciting the (3,0) cavity mode of a stack containing 580 junctions. In order to minimize self-heating, the THz source was mounted on a copper substrate using PbSn solder. We will discuss the choice of mesa dimensions and cavity mode, and strategies for the design of $Bi_2Sr_2CaCu_2O_8$ THz devices that are intended to operate at 77 Kelvin or above. We will also briefly discuss the as-fabricated sidewall angle of the stacks, and its implications for efficient synchronization of the stacked Josephson junctions.

**Assistant Professor Dr. Timothy Benseman**
6530 Kissena Boulevard
Queens, New York 11367, U.S.A.
Phone: +1 718-997-3369
Fax: +1 718-997-3349
E-mail: Timothy.Benseman@qc.cuny.edu
Web: https://physics.qc.cuny.edu/people/faculty/tbenseman

Vladimir M. Fomin

# Topological transitions in rolled-up superconductor nanomembranes under a strong transport current

*V. M. Fomin[1,2], R. O. Rezaev[1,3], E. I. Smirnova[1], O. G. Schmidt[1,4-6]*

[1]*Leibniz IFW Dresden, Institute for Integrative Nanosciences, Dresden, Germany*
[2]*Moldova State University, Laboratory of Physics and Engineering of Nanomaterials, Department of Theoretical Physics, Chisinau, Moldova*
[3]*Tomsk Polytechnic University, Tomsk, Russia*
[4]*TU Chemnitz, Material Systems for Nanoelectronics, Chemnitz, Germany*
[5]*TU Chemnitz, Research Center MAIN, Chemnitz, Germany*
[6]*TU Dresden, Nanophysics, Faculty of Physics, Dresden, Germany*

Topological defects such as vortices and phase slips in a superconductor system manifest spatial patterns and dynamics that are closely associated with the geometric design in curved micro- and nanostructures of superconductors [1]. We report on a topological transition between superconducting vortices and phase slips under a strong transport current in an open superconductor nanotube with a submicron-scale inhomogeneity of the normal-to-the-surface component of the applied magnetic field [2]. When the magnetic field is orthogonal to the axis of the nanotube, which carries the transport current in the azimuthal direction, the phase-slip regime is characterized by the vortex/antivortex lifetime $\sim 10^{-14}$ s versus the vortex lifetime $\sim 10^{-11}$ s for vortex chains in the half-tubes, and the induced voltage shows a pulse as a function of the magnetic field. This non-monotonous behavior is attributed to the occurrence of a phase-slip area. The topological transition between the vortex-chain and phase-slip regimes determines the magnetic-field–voltage and current–voltage characteristics of curved superconductor nanomembranes to pursue high-performance applications in advanced electronics (e.g., as novel superconductor switching-based detectors).

**Prof. Dr. Vladimir M. Fomin**
Helmholtzstraße 20
01069 Dresden, Germany
Phone: +49 351-4659-780
Fax: +49 351-4659-782
E-mail: v.fomin@ifw-dresden.de
Web: https://www.ifw-dresden.de/about-us/people/prof-dr-vladimir-fomin

## Spin waves and related dynamic phenomena in layered superconductors with helical magnetic structure

*A. E. Koshelev*

*Materials Science Division, Argonne National Laboratory, 9700 South Cass Ave., Lemont, Illinois, 60439, USA*

We evaluate the spin-wave spectrum and dynamic susceptibility in a layered superconductor with helical interlayer magnetic structure [1]. We especially focus on the structure in which the moments rotate 90° from layer to layer realized in the iron pnictide $RbEuFe_4As_4$. While in nonmagnetic superconductors low-frequency magnetic field decays on the distance of the order of the London penetration depth, spin waves mediate its propagation to much larger distances limited by external dissipation mechanisms. The long-range electromagnetic interactions between the oscillating magnetic moments strongly increase the frequency of the mode coupled with uniform field and this increase exists only within a narrow range of the c-axis wave vectors of the order of the inverse London penetration depth. The key feature of materials like $RbEu Fe_4As_4$ is that this uniform mode corresponds to the maximum frequency of the spin-wave spectrum with respect to c-axis wave vector. As a consequence, the high-frequency surface resistance acquires a very distinct asymmetric feature spreading between the bare and renormalized frequencies. We also consider excitation of spin waves via AC Josephson effect in a tunneling contact between helical-magnetic and conventional superconductors and explore the interplay between the spin-wave features and geometrical cavity resonances in the current-voltage characteristics.

This work was supported by the US Department of Energy, Office of Science, Basic Energy Sciences, Materials Sciences and Engineering Division.

**Dr. A.E.Koshelev**
Materials Science Division
Argonne National Laboratory
9700 South Cass Ave., Lemont, Illinois, 60439
USA
Phone: +1 630-252-9592
E-mail: koshelev@anl.gov
Web: https://www.anl.gov/profile/alexei-e-koshelev

## Using the database and information technologies in studying functional nanostructures

*Yu. B. Savva*

*Orel State University named after I.S. Turgenev, Orel, Russia*

When carrying out scientific research, it becomes necessary to save and subsequent computer processing of the results of experiments. For this purpose, the Research Laboratory of Functional Nanostructures has developed the database «NIL-FN». This database is a tool for helping in the daily activities of laboratory staff: storage of information about materials and equipment, development and description of experiments, data marking, storage and organization of results. All experimental data is stored and described scientifically significant in terms of target, method, and interpretation of results. To track the creation, changing or deleting any entries in the database using its management system, control logs are generated with login, date and user time stations.

For processing and analyzing experimental data, applications are used to the Python programming language using a widespread library with open source PANDAS. This is due to the fact that the main object of this library - DataFrame may contain inhomogeneous data types: a floating point number, integers, lines, dates, time, etc., which can be structured in the form of hierarchy and index, which is consistent with the adopted hierarchical attitude between Tables in which experimental data are stored. In the developed processing and data analysis applications, their visualization is provided, because It simplifies and speeds up the execution of data analysis. The main purpose of data visualization is the study (for example, the search for obvious patterns, emissions, etc.) and the presentation of the results in a visual form. With visual representation of the data (in the form of graphs, histograms or other forms), regularities are becoming obvious. To visualize the results of the experimental analysis in the NIL-FN database applications, the MATPLOTLIB library, designed to build 2D python graphs, which allows you to receive high-quality images to publish in various print copies and in interactive media on different platforms. Thus, users of NIL-FN database can easily build graphs, histograms, scattering diagrams and much more using Python applications with all multiple lines of the link code to the MATPLOTLIB library.

**Acknowledgements**

The study was carried out with the financial support of the Russian Scientific Fund, the project 20-62-47009 "Physical and engineering foundations of the calculators are not the background of Nymanovsk architecture based on superconductor spintronics."

**Prof. Assist. PhD in CS Yuri B. Savva**
95, Komsomolskaya st.
Orel region, Orel,
302026, Russia
Phone: +7 910-207-6185
Fax: +7 4862 751-318
E-mail: su_fio@mail.ru
Web: http://oreluniver.ru/employee/119

# Nonmonotonous temperature dependence of Shapiro steps in YBCO grain boundary junction

*L. S. Revin[1,2], D. V. Masterov[1], A. E. Parafin[1], S. A. Pavlov[1], A. L. Pankratov[1,2,3]*
*[1]Institute for Physics of Microstructures of RAS, GSP-105, Nizhny Novgorod, 603950, Russia*
*[2]Nizhny Novgorod State Technical University n.a. R.E. Alekseev, Nizhny Novgorod, Russia*
*[3]N.I. Lobachevsky State University of Nizhni Novgorod, Nizhny Novgorod, Russia*

In recent years, the limiting characteristics of detectors and mixers based on high-temperature superconducting Josephson junctions (JJs) have been actively studied. In the majority of works, an increase in sensitivity at low temperatures has been demonstrated, although a part of papers indicate the receiver's operation optimum at intermediate temperatures between the liquid nitrogen and helium temperatures. We report on a temperature dependence of the first Shapiro step amplitude for an external signal with frequencies of 72 GHz and 265 GHz acting on $YBa_2Cu_3O_{7-\delta}$ Josephson grain boundary junction [1-3]. The observed non-monotonous behavior of the step height in the limit of low signal power is discussed. The step heights are in agreement with the calculations based on the resistively-capacitively shunted junction model and Bessel theory. The occurrence of the receiving optima is explained by the mutual influence of the varying critical current and the characteristic frequency. The maximum response to a 72 GHz signal has an optimum at 70 K, to a 265 GHz signal - 50 K. The obtained optima arise at certain JJ parameters: normal resistance $R_N$, critical current $I_C(T)$, characteristic frequency $\omega_C(T)$. For specific purposes and operation regions, it is possible to tune JJ parameters to operate in the optimal regime.

The work is supported by the RSF (Project No. 20-79-10384).

**Dr. Leonid S. Revin**
603087 Nizhny Novgorod, Russia
Phone: +7 920-257-79-78
E-mail: rls@ipmras.ru
Web: http://ipmras.ru/en/structure/people/rls

Reinhold Kleiner

# Space-time crystalline order of a high-critical-temperature superconductor with intrinsic Josephson junctions

**Reinhold Kleiner[1*], Xianjing Zhou[2], Eric Dorsch[1], Xufeng Zhang[2], Dieter Koelle[1], and Dafei Jin[2]**
[1]*Physikalisches Institut, Center for Quantum Science (CQ) and LISA[+], Universität Tübingen, Auf der Morgenstelle 14, 72076 Tübingen, Germany*
[2]*Center for Nanoscale Materials, Argonne National Laboratory, Argonne, Illinois 60439, USA*
[*]*Corresponding author, E-mail address: Kleiner@uni-tuebingen.de*

We theoretically demonstrate that the high-critical-temperature superconductor $Bi_2Sr_2CaCu_2O_{8+x}$ (BSCCO) is a natural candidate for the recently envisioned classical space-time crystal. BSCCO intrinsically forms a stack of Josephson junctions. Under a periodic parametric modulation of the Josephson critical current density, the Josephson currents develop coupled space-time crystalline order, breaking the continuous translational symmetry in both space and time. The modulation frequency and amplitude span a (nonequilibrium) phase diagram for a so-defined spatiotemporal order parameter, which displays rigid pattern formation within a particular region of the phase diagram. Based on our calculations using representative material properties, we propose a laser-modulation experiment to realize the predicted space-time crystalline behavior. Our findings bring new insight into the nature of space-time crystals and, more generally, into nonequilibrium driven condensed matter systems.

## References

1. R. Kleiner et al., arXiv:2012.01387

**Prof.Dr. Reinhold Kleiner**
Physikalisches Institut,
Center for Quantum Science (CQ) and LISA[+],
Universität Tübingen,
Auf der Morgenstelle 14, 72076
Tübingen, Germany
E-mail: Kleiner@uni-tuebingen.de

# Exotic superconducting states in FeSe-based materials

*T. Shibauchi*

*Department of Advanced Materials Science, University of Tokyo, Kashiwa, Japan*

The interplay among magnetism, electronic nematicity, and superconductivity is the key issue in strongly correlated materials including iron-based, cuprate, and heavy-fermion superconductors. Magnetic fluctuations have been widely discussed as a pairing mechanism of unconventional superconductivity, but recent theory predicts that quantum fluctuations of electronic nematicity, which is characterized by rotational symmetry breaking, may also promote high-temperature superconductivity. FeSe-based superconductors are suitable to study this issue [1], because FeSe exhibits a nonmagnetic nematic order that can be suppressed by S or Te substitution for Se. I will review recent studies of FeSe-based superconductors, which show quite exotic superconducting states. In $FeSe_{1-x}S_x$ superconductors, the nematic order can be completely suppressed at $x$=0.17, above which the superconducting properties change drastically with a significantly reduced critical temperature $T_c$ [2,3]. From recent muon spin rotation (μSR) measurements [4], we find evidence for a novel ultranodal pair state with broken time reversal symmetry [5]. In the Te substitution case, however, we find quite different behavior; the suppression of nematic order leads to an enhancement of $T_c$, which is likely associated with quantum critical fluctuations of nematicity [6].

**Prof. Takasada Shibauchi**
5-1-5 Kashiwanoha, Kashiwa, Chiba 277-8561, Japan
Phone: +81 4-7136-3774
Fax: +81 4-7136-3774
E-mail: shibauchi@k.u-tokyo.ac.jp
Web: http://qpm.k.u-tokyo.ac.jp/shibauchi/index_e.html

# Steady Floquet-Andreev States Probed by Tunnelling Spectroscopy

*Department of Physics, POSTECH, Pohang 37673, South Korea*

Engineering quantum states through light-matter interaction has created a new paradigm in condensed matter physics. A representative example is the Floquet-Bloch state, which is generated by time-periodically driving the Bloch wavefunctions in crystals. Previous attempts to realise such states in condensed matter systems have been limited by the transient nature of the Floquet states produced by optical pulses, which masks the universal properties of non-equilibrium physics. Here, we report the generation of steady Floquet Andreev (F-A) states in graphene Josephson junctions by continuous microwave application and direct measurement of their spectra by superconducting tunnelling spectroscopy [1]. We present quantitative analysis of the spectral characteristics of the F-A states while varying the phase difference of superconductors, temperature, microwave frequency and power. The oscillations of the F-A state spectrum with phase difference agreed with our theoretical calculations. Moreover, we confirmed the steady nature of the F-A states by establishing a sum rule of tunnelling conductance, and analysed the spectral density of Floquet states depending on Floquet interaction strength. This study provides a basis for understanding and engineering non-equilibrium quantum states in nano-devices.

**Prof. Gil-Ho Lee**
Department of Physics, POSTECH, Pohang 37673, South Korea
E-mail: lghman@postech.ac.kr

# Josephson coupling and pair density waves in stripe-ordered superconductors

*Qiang Li[1,2]*
[1]*Department of Physics and Astronomy, Stony Brook University, USA*
[2]*Condensed Matter Physics and Materials Science Division, Brookhaven National Laboratory, USA*

Among cuprate high-temperature superconductors, spin-stripe order is seen most clearly in the $La_2CuO_4$-based family. This order, occurring together with charge-stripe order, is strongest for doped hole concentration, x, near 1/8. A prominent example is $La_{2-x}Ba_xCuO_4$ (LBCO) [1]. The onset of spin-stripe order is associated with the occurrence of two-dimensional superconductivity (2D SC). The depression of 3D SC order due to the frustration of the interlayer Josephson coupling is attributed to pair-density-wave (PDW) order [2].

Both Zn-doping and c-axis magnetic fields have been observed to increase the spin stripe order in $La_{2-x}Ba_xCuO_4$ with x close to 1/8. For x = 0.095, the applied magnetic field also causes superconducting layers to decouple, presumably by favoring pair-density-wave order that consequently frustrates interlayer Josephson coupling. Here we show that introducing 1% Zn also leads to an initial onset of two-dimensional (2D) superconductivity, followed by 3D superconductivity at lower temperatures, even in zero field [3]. The angle-resolved c-axis transport measurements of the $La_{2-x}Ba_xCuO_4$ with x = 1/8 revealed periodic dependence of resistivity and Josephson current on the direction of the in-plane magnetic field, which may be used for probing the structure of pair density wave in striped superconductors.

**Prof. Dr. Qiang Li**
Department of Physics and Astronomy, Stony Brook University
Stony Brook, NY 11794 3800 USA
Phone: (631)-344-4490
E-mail: qiang.li@stony.brook.edu or qiangli@bnl.gov
Web: https://you.stonybrook.edu/qiangli/

# The pulse-driven AC Josephson voltage standard

*O. Kieler, H. Tian, R. Gerdau, J. Kohlmann, R. Behr*
*Physikalisch-Technische Bundesanstalt (PTB), Braunschweig, Germany*

The pulse-driven AC Josephson voltage standard, often called "Josephson Arbitrary Waveform Synthesizer" (JAWS), is based on large series arrays of SNS (S…superconductor, N…normal metal) Josephson junctions with $Nb_xSi_{1-x}$ as barrier material. Arbitrary AC voltages can be synthesized with high spectral purity and quantum precision. Series arrays of up to 5-stacked SNS junctions [1] have been implemented with up to 60 000 junctions for a 10 mm x 10 mm chip. A sophisticated complex fabrication technology was developed to achieve a high device yield. Electron beam lithography, chemical mechanical polishing and atomic layer deposition are the key components of this technology. The arrays are operated at 4 K by short current pulses, delivered by a fast pulse pattern generator, with a maximum data rate of 30 Gbit/s. The desired output waveform is encoded by means of a higher-order sigma-delta modulation. Output voltages of more than 2 V were demonstrated at PTB by operating 16 arrays in series [2]. Recent efforts to further increase the output voltage are an optical pulse-drive [3] or an on- chip power divider [4]. Many applications in voltage metrology have been established already.

## Acknowledgement

This work was supported in part by the EMPIR Program co-financed by the Participating States and from the European Union's Horizon 2020 Research and Innovation Program (Joint Research Project 15SIB04 QuADC) and in part by the German Federal Ministry for Economic Affairs and Energy due to a decision of the German Bundestag under Project ZF4104104AB7.

**Dr. Oliver Kieler**
Physikalisch-Technische Bundesanstalt (PTB) Bundesallee 100
38116 Braunschweig, Germany
Phone: +49 531-592-2410
Fax: +49 531-592-692405
E-mail: oliver.kieler@ptb.de
Web: http://www.ptb.de/

**Mikhail Galin**

# Fine Structure of Radiation Spectra from Large Josephson Systems

*M. A. Galin[1,2], O. Kieler[3], M. Yu. Levichev[1], A. I. El'kina[1], V. V. Kurin[1]*
[1]*Institute for Physics of Microstructures RAS, 603950 Nizhny Novgorod, Russia*
[2]*Moscow Institute of Physics and Technology, 141700 Dolgoprudny, Russia*
[3]*Physikalisch-Technische Bundesanstalt, 38116 Braunschweig, Germany*

At present, many scientific groups are investigating superconducting systems containing Josephson junctions (JJs) as generators of subterahertz and terahertz radiation [1–4]. For the arrays with a large number of JJs the question of their coherent contribution to output radiation becomes the key point for practical applications. The synchronization is usually provided mainly by resonant modes which can be excited in cavities of very different types [4–6]. In this work we demonstrate that some peculiarities in external electrodynamical environment can affect the synchronization regimes in large JJ arrays that manifests in fine structure of spectral properties of their radiation.

We investigate transport (IVC) and radiation (spectrum, emission power) characteristics of the arrays containing up to 10 thousands Nb/NbSi/Nb junctions composed in different designs. Measurements were performed in liquid helium dewar at 4.2 K. Current steps in IVCs were associated with some internal resonances excited in parts of the arrays. Data of spectral measurements of the array with 9996 JJs are considered to indicate the effects of two-frequency Josephson emission and non-Josephson generation. The radiation power from this array detected by the room receivers reaches 14 μW at 255 GHz. However these data, also as the spectrum data and IVC, are very sensitive to manipulation with the measurement probe that we explained by the influence of external electrodynamical elements inside the probe to the array. All of these conclusions are confirmed by the results of numerical simulations of Josephson antennas combined the solutions of the Maxwell equations and of nonlinear equations for the JJs [7]. In more advanced current version of our software the parameter scattering of JJs has been introduced in the algorithm.

The work was supported by the Russian Science Foundation, grant No. 20-42-04415.

**Dr. Mikhail Galin**
Institute for Physics of Microstructures RAS, Academicheskaya Str.7,
603087 Afonino, Nizhny Novgorod region, Russia
Phone: +7 831-417-94-83
E-mail: galin@ipmras.ru
Web: http://ipmras.ru/en/structure/people/galin

Leonid S. Kuzmin

# Development of a Single Photon Counter for a Yoctojoule Energy Range

*L.S. Kuzmin[1,3], A.L. Pankratov[1,2], L.S. Revin[1,2], A.V. Gordeeva[1,2,3], A.A. Yablokov[1,2], E.V. Il'ichev[4]*
[1]*Nizhny Novgorod State Technical University n.a. R. E. Alekseev, Nizhny Novgorod, Russia*
[2]*Institute for Physics of Microstructures of RAS, Nizhny Novgorod, Russia*
[3]*Chalmers University of Technology, Gothenburg, Sweden*
[4]*Leibniz Institute of Photonic Technology, Jena, Germany*

We have obtained convincing experimental evidence of switching aluminum Josephson Single Photon Counter (SPC) to the resistive state initiated by absorption of a few photons at 10 GHz. The corresponding photon energy of only 7 yoctojoule is smaller by 4 orders than for infrared photons. The switching of SPC by a strongly attenuated harmonic signal (represented by a Poissonian distribution of photons) has been detected in a single-click regime at the temperature of 50 mK. By comparison with Poissonian statistics, it is shown that the switching probability corresponds to the detection of 1, 2, 3, 4, and 5 photons. The maximal power of the absorbed signal is estimated by the photon-assisted tunnelling steps using Tien–Gordon theory as 70 fW and is further attenuated into a few fW range. Using the sub-µm Al tunnel junction with 23 nA critical current, we demonstrate a breakthrough in the detector performance utilizing the phase diffusion regime, allowing to increase the dark count time dramatically. The suggested Single Photon Counter prototype can be further optimized to match the requirements for axion search.

We apply the same methodology as has been used before in G. N. Gol'tsman et al., Appl. Phys. Lett., **79**, 705 (2001) for detecting infrared photons, where the number of detected photons is extracted by comparison with Poissonian statistics. We, therefore, for the first time, realize a single-click microwave photon counter with extremely low photon energy down to 7 yoctojoule.

This work was supported by the Russian Science Foundation (Project No. 19-79-10170).

**Prof. Dr. Leonid S. Kuzmin**
Principal researcher,
Laboratory of Superconducting Nanoelectronics
of Nizhny Novgorod State Technical University
Chalmers University of Technology, Gothenburg, Sweden
Phone: +46768985404
E-mail: kuzmin@chalmers.se
Web: https://www.nntu.ru/phonebook/card/spravochnik/laboratoriya-sverhprovodnikovoy-nanoelektroniki/kuzmin-leonid-sergeevich

# Multiband electronic structure as a key to high temperature superconductivity and new quantum applications

**Prof. Alexander Kordyuk**
Member of National Academy of Sciences of Ucraine,
Director of Kyiv Academic University
E-mail: kordyuk@gmail.com

# Magnetic Field-Induced "Mirage" Gaps and Triplet Josephson Effect in Ising Superconductors

**Prof.Dr. Wolfgang Belzig**
Tel: +49-7531-88-4782
Fax: +49-7531-88-3090
E-mail: wolfgang.belzig@uni-konstanz.de
Postal Address:
Universität Konstanz
Fachbereich Physik
Fach 703
D-78457 Konstanz

## Modeling of technological processes for the formation of nanostructures on a solid surface

### *A.V. Vakhrushev*[1,2]

*[1]Udmurt Federal Research Center of the Ural Branch of the Russian Academy of Sciences, Izhevsk, Russia*

*[2]Kalashnikov Izhevsk State Technical University, Izhevsk, Russia*

Forming nanostructures on the solids surface is one of the promising nanotechnological processes. It has been established that changes in the atomic structure of the solid surface due to the formation of nanostructures result both in a significant change in various physical properties of the surface, and in an increase in its durability, strength, hardness, wear resistance and defects on the surface healing. There are many different methods for forming nanostructures on solid surfaces: surface modification with nano-elements (nanoparticles, fullerenes and fullerites, graphene and nanotubes), formation of a nanocomposite layer on the surface, forming quantum dots and whiskers on the surface, implantation of ions into the solid surface, laser surface treatment and other processes. The above processes are very complex and for their optimization require detailed research both by experimental and theoretical methods of mathematical modeling [1-2]. The paper provides a comparative review of different methods of forming nanostructures on the solids surface and various aspects of mathematical modeling of these processes.

### Acknowledgements

The study was financially supported by the Russian Science Foundation Grant (RSF) Nr. 20-62-47009.

**Prof. Dr. Alexander V. Vakhrushev**
Udmurt Federal Research Center of the Ural Branch of the Russian Academy of Sciences,
Baramzinoy 34,
426067 Izhevsk, Russia
Phone: +7 912-466-8029
Fax: + 7 3412- 50-79-59
E-mail: Vakhrushev-a@yandex.ru
Web: http://iam.udman.ru/ru/user/vakhrouchev-av

**Yury M. Shukrinov**

# Phase dynamics, IV-characteristics and magnetization dynamics of the $\boldsymbol{\varphi}_0$ Josephson junction

*Yu. M. Shukrinov[1,2], I. Rahmonov[1], A. Janalizadeh[3] , M. R. Kolahchi[3]*

[1]*BLTP, Joint Institute for Nuclear Research, Dubna, Moscow Region, 141980, Russia*
[2]*Dubna State University, Dubna, 141980, Russia*
[3]*Department of Physics, Institute for Advanced Studies in Basic Sciences, Zanjan 45195-1159, Iran*

The SFS $\boldsymbol{\varphi}_0$ Josephson junctions (JJ) with the phase shift proportional to the magnetic moment of ferromagnetic layer determined by the parameter of spin-orbit interaction, demonstrate a number of unique features important for superconducting spintronics and modern information technology [1,2]. The phase shift $\boldsymbol{\varphi}_0$ allows one to manipulate the internal magnetic moment using the Josephson current and the reverse phenomenon which leads to the appearance of the DC component in the superconducting current [3-5].

We demonstrate an anomalous features of the ferromagnetic resonance with an increase of the Gilbert damping. We found that the resonance curves demonstrate features of Duffing oscillator, reflecting the nonlinear nature of Landau-Lifshitz-Gilbert equation. The damped precession of the magnetic moment is dynamically driven by the Josephson supercurrent, and the resonance behavior is given by a Duffing spring. The resonance methods for the determination of spin-orbit interaction in the $\boldsymbol{\varphi}_0$ junction are proposed.

## Acknowledgements

The work was partially funded by the RFBR, according to the project 20-37-70056. Numerical calculations have been done within the framework of the project 18-71-10095 of the Russian National Fund.

**Prof. Dr. Yury M. Shukrinov**
Bogoliubov Laboratory of Theoretical Physics,
Joint Institute for Nuclear Research,
Dubna, Moscow Region, 141980, Russia
Phone: +7 4962163844 (off), +7 9150442981(mobile1)
+7 9771795931(mobile2), Fax: +7 4962165084
E-mail: shukrinv@theor.jinr.ru , yurishukrinov@yahoo.com

# Topological Order in HighTemperature Superconductors

**Prof. Fiodor Kusmartsev**
Loughborough (UK) and Khalifa ( Abu Dhabi, UAE) Universities
E-mail: f.kusmartsev@lboro.ac.uk

**Tairzhan Karabassov**

## Reentrant superconductivity in proximity to a topological insulator

_T. Karabassov[1], A. A. Golubov[2,3], V. M. Silkin[4,5,6], V. S. Stolyarov[3,7], A. S. Vasenko[1,8]_

[1]_HSE University, Moscow, Russia_

[2]_Faculty of Science and Technology and MESA Institute for Nanotechnology, University of Twente, Enschede, The Netherlands_

[3]_Moscow Institute of Physics and Technology, Dolgoprudny, Russia_

[4]_Donostia International Physics Center (DIPC), Paseo Manuel de Lardizabal 4, San Sebastian Donostia, Basque Country, Spain_

[5]_Departamento de Fisica de Materiales, Facultad de Ciencias Quimicas, UPV/EHU, San Sebastian, Basque Country, Spain_

[6]_IKERBASQUE, Basque Foundation for Science, Bilbao, Spain_

[7]_Dukhov Research Institute of Automatics (VNIIA), Moscow, Russia_

Superconducting hybrid structures with topological order and induced magnetization offer a promising way to realize fault-tolerant quantum computation [1]. However, the effect of the interplay between magnetization and the property of the topological insulator surface, otherwise known as spin-momentum locking on the superconducting proximity effect, still remains to be investigated. We relied on the quasiclassical self-consistent approach to consider the superconducting transition temperature in the two-dimensional superconductor/topological insulator (S/TI) junction with an in-plane helical magnetization on the TI surface [2]. It has emerged that the presence of the helical magnetization leads to the nonmonotonic dependence of the critical temperature on the TI thickness for both cases when the magnetization evolves along or perpendicular to the interface. The results obtained can be helpful for designing novel superconducting nanodevices and better understanding the nature of superconductivity in S/TI systems with nonuniform magnetization.

**PhD student, Tairzhan Karabassov**
34 Tallinskaya Ulitsa,
123458, Moscow, Russia
Fax: +7 (495) 916-88-29
E-mail: tkarabasov@hse.ru
Web: https://www.hse.ru/staff/Tair

**Valeriy Ryazanov**

# Superconducting phase inversions in mesoscopic superconductor-normal metal-ferromagnet Josephson structures controlled by nonequilibrium injections

*V. V. Ryazanov, T. E. Golikova, M. J. Wolf, D. Beckmann, G. A. Penzyakov, I. E. Batova, I. V. Bobkova, A. M. Bobkov*

We have studied experimentally Josephson effect in the crosslike S–N/F–S junctions driven in nonequilibrium by applying an injection current across the complex N/F weak link. The superconducting critical current upon increasing the injection current manifests the nonmonotonic behavior with two pronounced dips, which are supposed to be due to the double 0–π transition. The model taking into account the supercurrent-carrying density of states (SCDOS) of a S–N/F–S structure and spin injection into the N layer is developed to describe the observed effect.

**Prof. Valeriy Ryazanov**
Institute of Solid State Physics, Ryssian Academy of Sciences
Laboratory for Superconductivity
Russia
E-mail: valery.ryazanov@gmail.com

# Superconducting neuron for networks based on radial basis functions

**Prof. Klenov Nikolai**
M. Lomonosov Moscow State University, Moscow, Russia
E-mail: nvklenov@gmail.com
Tel: +9175378893

# Controllable two- *vs* three-state magnetization switching in single-layer epitaxial $Pd_{1-x}Fe_x$ films and $Pd_{0.92}Fe_{0.08}$/Ag/$Pd_{0.94}Fe_{0.04}$ heterostructure

*I. V. Yanilkin[1], A. I. Gumarov[1], G. F. Gizzatullina[1], R. V. Yusupov[1], L. R. Tagirov[2,3]*

[1]*Institute of Physics, Kazan Federal University, Kazan, Russia*
[2]*Zavoisky Physical-Technical Institute, FRC Kazan Scientific Center of RAS, Kazan, Russia*
[3]*Tatarstan Academy of Sciences, Institute of Applied Research, Kazan, Russia*

The generation of the long-range triplet component of the superconducting pairing at noncollinear orientations of magnetizations in ferromagnetic layered systems is extensively studied in magnetic Josephson junctions (MJJ) [1,2] and superconductive spin valves (SSV) [3]. Palladium-iron ($Pd_{1-x}Fe_x$) alloy with $x < 0.10$ has strong practical interest for such MJJ and SSV structures as a material for weak ferromagnetic layers with tunable magnetic properties [4]. Therefore, it is useful to investigate the switching properties of $Pd_{1-x}Fe_x$ films and heterostructures with the aim to create controllable non-collinear magnetic states.

Single epitaxial film $Pd_{0.92}Fe_{0.08}$(20 nm) (hereinafter PdFe1) and epitaxial heterostructure $Pd_{0.92}Fe_{0.08}$(20 nm)/Ag(20 nm)/$Pd_{0.94}Fe_{0.04}$(20 nm) (hereinafter PdFe1/Ag/PdFe2) were grown using molecular beam deposition technique [5]. Detailed measurements of the magnetoresistance have shown that the PdFe1 epitaxial film, being an easy-plane ferromagnet with a pronounced in-plane anisotropy, undergoes magnetization switching between two (with collinear magnetization directions) or three (including orthogonal to the two directions noted above) single-domain states, depending on the direction of the applied magnetic field. This unconventional magnetization switching is a consequence of a combination of the fourfold magnetocrystalline anisotropy and additional uniaxial magnetic anisotropy in the film plane.

The use of two magnetic layers PdFe1 and PdFe2 with different coercive fields, separated by a nonmagnetic spacer, makes it possible to obtain parallel, orthogonal, and antiparallel configurations of magnetic moments. It has been experimentally demonstrated that the PdFe1/Ag/PdFe2 heterostructure can be switched between three stable magnetic configurations in the film plane by rotating the magnetic moment of the soft magnetic layer with respect to the hard magnetic layer. Such a structure can serve as a magnetic actuator for switching an MJJ from the singlet conducting mode to the triplet conducting mode and vice versa.

**Dr. Igor V. Yanilkin**
Kremlevskaya18
420008 Kazan, Russia
Phone: +7 843 2337763
E-mail: yanilkin-igor@yandex.ru
Web: https://kpfu.ru/igor.yanilkin?p_lang=2

Anatolie Sidorenko

# Nanostructures Superconductor/Ferromagnet for Superconducting Spintronics

**_Anatolie Sidorenko[1,2], Roman Morari[1,2], Vladimir Boian[1], Evgeni Antropov[1], Andrei Prepelitsa[1], Yurii Savva[2], Nikolai Klenov[3], Igor Soloviev[3], Alexander Vakhrushev[2,4]_**

[1]Institute of Electronic Engineering and Nanotechnologies, Academiei str., 3/3, MD-2028, Chisinau, Republic of Moldova

[2]I.S. Turgenev Orel State University Komsomolskaya str. 95, 302026, Orel, Russia

[3]Lomonosov Moscow State University Skobeltsyn Institute of Nuclear Physics, Moscow, 119991, Russia

[4]Nanotechnology and Microsystems Department, Kalashnikov Izhevsk State Technical University, Studencheskaya 7, Izhevsk 426069, Russia

Layered superconductor-ferromagnet (S/F) nanostructures became an important base of superconducting spintronics. Theory of S/F hybrid heterostructures with two and more ferromagnetic layers predicts generation of a non-uniform superconductivity, a long-range odd-in-frequency triplet pairing at non-collinear alignment (NCA) of the F-layers magnetizations. Using the ideas of the superconducting triplet spin-valve we have fabricated functional nanostructures $Co/CoO_x/Cu_{41}Ni_{59}/Nb/$ were triplet pairing with switching from normal to superconducting state takes place. The resistance of the samples as a function of an external magnetic field shows that the system is superconducting at the collinear alignment of the $Cu_{41}Ni_{59}$ and Co layers magnetic moments, but switches to the normal conducting state at the NCA configuration. Upon cycling the in-plane magnetic field in the range and keeping temperature close to the superconducting transition, a memory effect has been detected, The fabricated S/F functional nanostructures can serve as the rapid operating switching element of superconducting electronics and memory element MRAM for novel computers generation.

## Acknowledgements

The study was financially supported by the Russian Science Foundation Grant ( RSF ) Nr. 20-62-47009 "Physical and engineering basis of computers non-von Neumann architecture based on superconducting spintronics" (modelling of the vacuum process, fabrication of the layered nanostructures) and project "SPINTECH" of the HORIZON-2020 program under GA No 810144 (low temperatures measurements).

**Prof. Dr. Anatolie Sidorenko**
Institute of Electronic Engineering and Nanotechnologies "D. GHITU"
Academiei 3/3, Kishinev MD2028 Moldova
Tel +37322-727072; FAX +37322-727088
E-mail: anatoli.sidorenko@kit.edu
Web: http://nanotech.md
ORCID 0000-0001-7433-4140
Linked In:  https://www.linkedin.com/in/anatoli-sidorenko-102a1629/

# Approaches to scaling superconducting digital circuits

**Prof. Soloviev Igor**
M. Lomonosov Moscow State University, Moscow, Russia
E-mail: igor.soloviev@gmail.com
Tel: +79262676589

**Irina V. Bobkova**

# Triplet superconductivity induced by moving condensate

_I. V. Bobkova[1,2,3], A. M. Bobkov[1], M. A. Silaev[4,2], A. A. Mazanik[2,5]_
[1]_Institute of Solid State Physics, Chernogolovka, Russia_
[2]_Moscow Institute of Physics and Technology, Dolgoprudny, Russia_
[3]_National Research University Higher School of Economics, Moscow, Russia_
[4]_Department of Physics and Nanoscience Center, University of Jyvaskyla, Finland_
[5]_BLTP, Joint Institute for Nuclear Research, Dubna, Russia_

It has been commonly accepted that electromagnetic fields suppress superconductivity by inducing the ordered motion of Cooper pairs. We demonstrate a mechanism which instead provides generation of superconducting correlations by moving the superconducting condensate. This effect arises in superconductor/ferromagnet heterostructures in the presence of Rashba spin-orbit coupling or extrinsic impurity-induced spin-orbit coupling. We predict the odd-frequency spin-triplet superconducting correlations called the Berezinskii order to be switched on in ferromagnets at large distances from the superconductor/ferromagnet interface by application of a static magnetic field or irradiation inducing condensate motion in the direction perpendicular to structural anisotropy axis, what opens great perspectives for low-dissipative spintronics [1,2]. The effect is shown to result in the unusual behaviour of Josephson current under the applied magnetic field, supercurrent-controllable 0-π transitions and photo-induced Josephson current.

**Dr. Irina V. Bobkova**
Institute of Solid State Physics,
Chernogolovka, Russia
E-mail: bobkova@issp.ac.ru

# Chirality of Bloch domain walls in exchange biased CoO/Co bilayer seen by waveguide-enhanced neutron spin-flip scattering

*Yu. Khaydukov, D. Lenk, V. Zdravkov, R. Morari, T. Keller, A. S. Sidorenko, L. R. Tagirov, R. Tidecks, S. Horn, B. Keimer*

Magnetic state of exchanged biased CoO(20nm)/Co(dF) bilayer (dF=5-20nm) was studied by means of polarized neutron reflectometry. By spacing of CoO/Co bilayer and Al2O3 substrate with Nb(20nm) layer we created waveguide structure which allowed us to significantly enhance intensity of spin-flip (SF) scattering in the position of optical resonances. For the trained sample with thinnest Co(5nm) we detected strong SF scattering at the resonance position (up to 30\% of incoming intensity) speaking about high non-collinearity of the system. As dF increases, the intensity of SF scattering linearly decreases. At the same time we observed asymmetry of up-down and down-up scattering channels at the resonance positions. We attribute this asymmetry to the Zeeman splitting of neutrons energies with different initial polarization taking place in high external field. Analysis, however, shows that the applied in the PNR experiment external field is not enough to quantitatively explain the observed asymmetry for the samples with dF> 5nm and we have to postulate presence of additional magnetic field produced by sample. We attribute this additional field to the stray field produced by chiral Bloch domain walls. The chirality of the domain walls can be explained by Dzyaloshinskii-Moriya interaction arising at the CoO/Co interface. Our results can be useful for designing of spintronic devices using exchange bias effect.

**Dr. Yury Khaydukov**
Max-Planck Institute for Solid State Research
ZWE FRM-II/NREX
Lichtenbergstr. 1
D-85747 Garching bei Muenchen
Germany
Phone: +49-89-289-14769
Fax: +49-89-289-14911
E-mail: y.khaydukov@fkf.mpg.de

**POSTER SESSION**

Roger Cattaneo

# Observation of a gradual collective excitation of a geometrical resonance in an array of Nb/NbSi/Nb Josephson junctions with increasing number of synchronized junctions

R.Cattaneo [1], O. Kieler [2], M. A. Galin [3,4], A. M. Klushin [3], V. M. Krasnov [1,4]

[1] Stockholm University, Physics Department, SE-10691 Stockholm, Sweden

[2] Physikalisch-Technische Bundesanstalt, 38116 Braunschweig, Germany

[3] Institute for Physics of Microstructures RAS, 603950 Nizhny Novgorod, Russia

[4] Moscow Institute of Physics and Technology, 141700 Dolgoprudny, Russia

Synchronization of many Josephson junctions (JJs) is required for achieving significant emission power from Josephson oscillators [1]. One of the most promising strategies of synchronization is via collective excitation of a geometrical resonance either inside [1] or outside [2] the junction. Successful synchronization of up to 9000 JJ's coupled to an external cavity mode, accompanied by a coherent superradiant emission, has been demonstrated in a 1D junction array [3]. However, dynamics of such synchronization is not fully understood: it depends strongly on the array geometry [3], a specific mode of the geometrical resonance [2,3] and many other factors [1].

Here we study experimentally a 1D array of Nb/NbSi/Nb JJ's, consisting of 7 meander lines with 1000 JJs in each line, similar to those studied in Refs. [2,3]. The JJs exhibit a hysteresis in current-voltage (*I-V*) characteristics, presumably of thermal origin. Presence of a small inhomogeneity (specifically introduced by applying a small magnetic field) together with the hysteresis, allows one-by-one switching of JJs from the superconducting to the resistive state. We observe that above a certain number of JJs, $N \gtrsim 100$, a resonant step starts to develop in the *I-V*. As shown in Ref. [3] the step is caused by a collective excitation of a cavity mode in the array electrode. The step amplitude gradually increases with increasing *N*, which is accompanied by a parabolic-like enhancement of the emission power, measured with an external bolometer. Our observation provides a clear evidence for the mutual feedback between the cavity mode and the oscillating JJs. Such the feedback provides an effective mechanism for array synchronization and coherent emission.

## Acknowledgements

The work was supported by the Russian Science Foundation Grants 19-19-00594 (experiment) and 20-42-04415 (sample fabrication).

Roger Cattaneo
E-mail: roger.cattaneo@fysik.su.se
Phone: +39 3341718323

# Broadband detector based on series YBCO grain boundary Josephson junctions

*E. I. Glushkov[1], A. V. Chiginev[1,2], L. S. Kuzmin[2,3], L. S. Revin[1,2]*

[1]*Institute for Physics of Microstructures of RAS, GSP-105, Nizhny Novgorod, 603950, Russia*
[2]*Nizhny Novgorod State Technical University n.a. R.E. Alekseev, Nizhny Novgorod, Russia*
[3]*Chalmers University of Technology, 41296, Gothenburg, Sweden*

High-temperature superconducting Josephson junctions (HTSC JJs) have great potential as promising materials for creating high-frequency devices such as microwave generators, sensitive detectors or mixers and voltage standard [1,2]. Due to the large energy gap of HTSC, the working frequency range of such devices has expanded to the THz frequency range [3]. However, for some applications the performances of the HTSC devices are limited by fairly low impedance of the JJ. One possible way to solve this problem is to replace a single JJ by a chain or an array of JJs combined with planar coupling structures.

We report on a modeling of a broadband receiving system based on a meander series of YBaCuO grain boundary junctions integrated into a log-periodic antenna. The technique for calculating the receiving properties of JJs in the direct detection regime was developed and applied. At the first stage, electromagnetic modeling of several geometries of antennas was carried out for effective receiving in the frequency range 50 - 800 GHz. The electromagnetic properties of the systems are investigated, namely the amplitude-frequency characteristic and the beam pattern. For the chosen antenna geometry, the cases of a series of junctions integrated as a receiving element were considered; the fraction of the absorbed power in each JJ is investigated and it is shown that the use of 7 junctions makes it possible to increase the total absorbed power at a fixed frequency of 250 GHz by a factor of 2.3. At the second stage, by solving a system of second-order differential equations related through the equation for the antenna, the characteristics of the detector at bias current regime were calculated: current-voltage characteristic, operating point, responsivity, noise and noise equivalent power (NEP). It is shown that the use of a series chain of junctions allows to improve the responsivity by a factor of 2.5, NEP value by a factor of 1.5, and the dynamic range by a factor of 6-8. As a result, for the standard technology of YBaCuO magnetron sputtering on a bicrystal substrate for a temperature of 77 K, NEP in the region of $3 \cdot 10^{-13}$ W/$\sqrt{Hz}$ is obtained.

The work is supported by the RSF (Project No. 20-79-10384).

**Dr. Alexander V. Chiginev**
603087 Nizhny Novgorod, Russia
Phone: +7 910-877-41-58
E-mail: chig@ipmras.ru
Web: http://ipmras.ru/en/structure/people/chig

# Numerical simulation of thermal and radiative properties of THz sources based on Bi$_2$Sr$_2$CaCu$_2$O$_{8+\delta}$ mesa structures

*M. M. Krasnov*[*1,2], *N.D. Novikova*[1,2], *R. Cattaneo*[3], *V. M. Krasnov*[3,2]
[1]*Keldysh Institute of Applied Mathematics of RAS, Moscow, Russia*
[2]*Moscow Institute of Physics and Technology, Dolgoprudny, Russia*
[3]*Stockholm University, Department of Physics, Stockholm, Sweden*

Impedance matching and self-heating are limiting factors for performance of THz sources based on **Bi$_2$Sr$_2$CaCu$_2$O$_{8+\delta}$** (Bi-2212) intrinsic Josephson junctions (IJJ's) [1]. Here we present numerical simulation of thermal and radiative properties of such devices with different structure and geometry. It is demonstrated that electrodes are playing an important role. First, they strongly contribute to heat removal from the mesa, as shown in Fig. 1, and, thus, help in reduction of self-heating. Second, electrodes are acting as microwave antennae and help in impedance matching between the mesa and the open space [2,3], thus, enhancing effectiveness of electromagnetic wave emission. We test various device geometries and conclude that devices based on Bi-2212 whiskers may provide significant advantages because both the electrodes and the whisker itself work as a matching turnstile-type antenna, facilitating significant emission, as shown in Fig. 2. These conclusions are in agreement with recent experimental data on whisker-based devices [4].

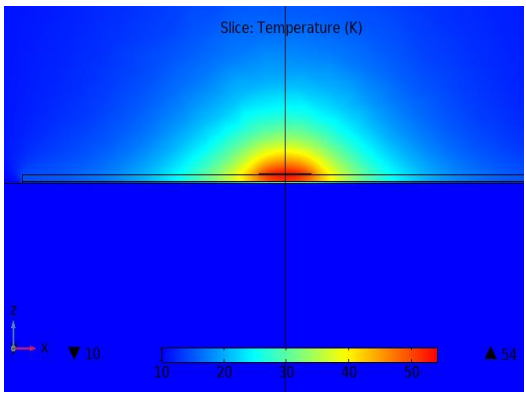
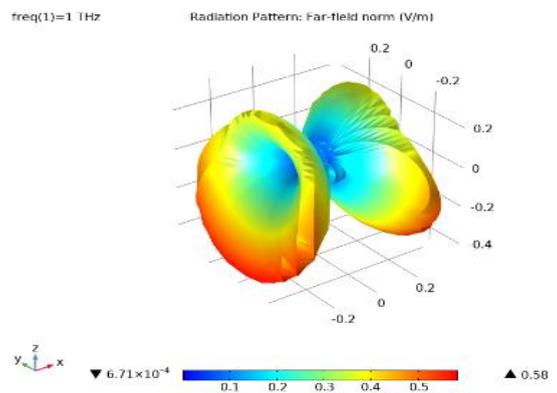

Fig. 1. Numerical simulation of temperature distribution in the case when heat removal occurs predominantly via heat-exchange He4 gas. In this case the top electrode acts as a cooling fin and makes the main contribution to heat removal. This is seen from the simple spherical temperature distribution above the electrode

Fig. 2. Numerical simulation of radiation emission pattern at f = 1 THs from a Bi-2212 whisker-based device with a cross-dipole (turnstile) type geometry. The directionality of emission is following the direction of the whisker. This demonstrates antenna-like operation of the whisker (and the electrodes)

## Acknowledgements
The work was supported by the Russian Science Foundation, Grant 19-19-00594. The work was accomplished during a sabbatical period of V.M.K. at MIPT, supported by the Faculty of Science at SU.

**Dr. Mihail Krasnov**
Keldysh Institute of Applied Mathematics
Russia
E-mail: kmm@kiam.ru

**Kirill Kulikov**

# Influence of external radiation on josephson junction + nanomagnet system

_K. V. Kulikov [1], M. Nashaat [1,2], M. Sameh [2], D. V. Anghel [3], A. T. Preda [3,4], and Yu. M. Shukrinov [1,4]_
[1]BLTP, JINR, Dubna, Moscow region, 141980, Russia
[2]Department of Physics, Faculty of Science, Cairo University, 12613, Giza, Egypt
[3]Horia Hulubei National Institute for R& D in Physics and Nuclear Engineering, Măgurele, Romania
[4]University of Bucharest, Faculty of Physics, Bucharest, Romania
[5]Dubna State University, Dubna, Russia

We investigate Kapitsa-like pendulum effects in the magnetic moment dynamics of a nanomagnet coupled to a Josephson junction under external periodic drive. Generated by the Josephson junction and external drive magnetic field play the role of the oscillating force of the suspension point in analogy with the Kapitsa pendulum. The high frequency oscillations change the position of stability of magnetic moment. The magnetic field of the quasiparticle current of the Josephson junction determines the frequency dependence of the magnetic moment's stable position. We obtain simple analytical formulas for the stable position of magnetic system both under external periodic drive and without it. The influence of external periodic drive on the voltage value of complete reorientation have been demonstrated.

**Kirill Kulikov**
Bogoliubov Laboratory of Theoretical Physics (BLTP)
Joint Institute for Nuclear Research
Dubna, Moscow region, 141980, Russia
E-mail: kirknemec2@mail.ru; kulikov@theor.jinr.ru

# Development of Dichroic 220/240 GHz Parallel Arrays of Cold-Electron Bolometers with Slot Antennas for LSPE project

**_D.A. Pimanov[1,3], L.S. Kuzmin[1,3], A.V. Chiginev[1,2], A.L. Pankratov[1,2], L.S. Revin[1,2]_**

[1]_Nizhny Novgorod State Technical University n.a. R. E. Alekseev, Nizhny Novgorod, Russia_
[2]_Institute for Physics of Microstructures of RAS, Nizhny Novgorod, Russia_
[3]_Chalmers University of Technology, Gothenburg, Sweden_

The samples of dichroic 220/240 GHz channels for the Large-Scale Polarization Explorer (LSPE) mission [1] were fabricated in Chalmers University of Technology. These samples consist of two parallel arrays of 32 Cold-Electron Bolometers (CEBs) [2-3] and slot antennas as on-chip filters for frequencies 220 and 240 GHz based on a 7*7 mm chip with substrate thickness of about 280 um. Measurements with black body power load are in progress in the dilution refrigerator at 50 to 300 mK temperatures. Measurements with YBCO-based oscillator [4] in cryostat are planned to investigate the response of these arrays in the frequency range from 200 to 300 GHz.

Electron cooling from 300 to 150 mK was achieved for these samples. The current/power responsivity of the arrays is about $2*10^4$ A/W at phonon temperature 300 mK. The obtained parameters and the shortcomings of the investigated samples are analyzed to improve the noise-equivalent power.

This work was supported by the Russian Science Foundation (Project No. 21-79-20227).

**Dmitry A. Pimanov**
PhD student,
Laboratory of Superconducting Nanoelectronics
of Nizhny Novgorod State Technical University
Phone: +7 9875543604
E-mail: macpimanov@gmail.com
Web:   https://www.nntu.ru/phonebook/card/spravochnik/laboratoriya-sverhprovodnikovoy-nanoelektroniki/pimanov-dmitrii-alekseevich

**Vladimir Boian**

## Pregătirea și investigarea Joncțiunii JOSEPHSON Nb/NiPt/Nb

*D. GHITU Institutul de Inginerie Electronica si Nanotehnologii, Chisinau, Moldova*

Funcționarea electronicii Josephson necesită de obicei determinarea curentului critic Josephson – Ic , care este afectat atât de diferite fluctuații și de zgomotele de măsurare. În acest experiment arătăm că Ic poate fi extras cu precizie utilizând prima și a treia armonică măsurători de blocare a rezistenței joncțiunii [1]. Măsurătoriile în experiment sau efectuat asupra probelor de tipul S – F – S : Nb – PtNi – Nb. În figura 1 de mai jos anexez caracteristica volt – amperică a joncțiunii în absența cîmpului la temperatura T = 4.47K:

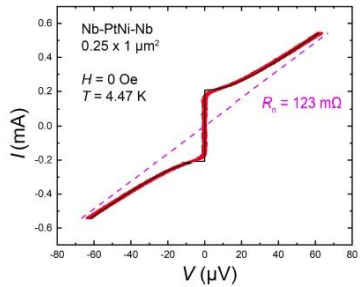

Fig. 1

Caracteristicile acestei probe în cimp magnetic se expun mai jos:

În Fig.2a) avem caracteristica I – V a probei la 4.47K cînd i se aplică probei paralel cîmpurile : 0, 25, 50, 100 Oe; imaginea b) și c) modulația de cîmp a primei și a treia garmonică a rezistenței probei în cîmp magnetic și d) modulația câmpului magnetic măsurată experimental în Ic(exp) – punctele negre, și curenți critici reconstituiți din prima armonică de și a 1-a (roșie) și a 3-a (linia albastră) de blocare a rezistenței probei.

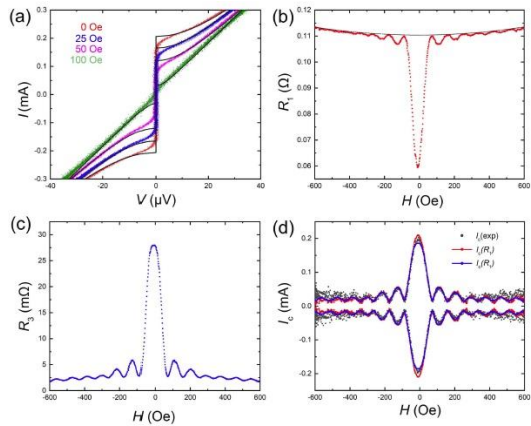

Fig. 2

### References
Accurate determination of the Josephson critical current by lock-in measurements.
https://doi.org/10.3390/nano11082058

Lucrarea efectuat în cadrul proiectului HORIZON-2020, SPINTECH, grant agreement Nr. 810144.

**Vladimir Boian**
PhD student,
D. GHITU Institutul de Inginerie Electronica si Nanotehnologii
Chisinau, Moldova
E-mail: boianvladimir@gmail.com

# Development of a fabrication process for deep submicron SQUID circuits with three independent Nb layers

*S. Wolter*[1], *J. Linek*[2], *J. Altmann*[3], *T. Weimann*[1], *S. Bechstein*[3], *R. Kleiner*[2], *J. Beyer*[3], *D. Koelle*[2] *and O. Kieler*[1]

[1] *Physikalisch-Technische Bundesanstalt (PTB), Fachbereich Quantenelektronik, Braunschweig, Germany*

[2] *Universität Tübingen, Physikalisches Institut, Center for Quantum Science (CQ) and LISA$^+$, Germany*

[3] *Physikalisch-Technische Bundesanstalt (PTB), Fachbereich Kryosensorik, Berlin, Germany*

To increase the integration density and the complexity of structures on-chip, we developed an extension of our established thin-film planar fabrication process for sub-µm-scale superconducting quantum interference devices (nanoSQUIDs) from originally two to three independent Nb layers. The process is based on the combination of electron beam lithography with chemical-mechanical polishing. Overdamped SNS (S: superconductor, N: normal conductor) Nb-HfTi-Nb Josephson junctions with sizes down to 80 nm x 80 nm for dc nanoSQUIDs having 100-nm-dimensions for Nb lines are realized already. Due to this miniaturization of their features, nanoSQUIDs offer high spatial resolution und high spin sensitivity approaching the order of 1 µB/Hz$^{1/2}$ (µB is the Bohr magneton), making them suitable for the detection of the magnetic states of magnetic nanoparticles. With the extension of the fabrication process the integration density can be increased further and superconducting vias to all Nb layers can be realized without the normal conducting HfTi barrier. We present results on the yield of this process and measurements of SQUID transport and noise properties.

**Acknowledgements**

This work was supported by the DFG (KI 698/3-2, KO 1303/13-2, BE 6680/1-2).

**Dr. Silke Wolter**
Bundesallee 100
38116 Braunschweig, Germany
Phone: +49 531 592 2465
E-mail: silke.wolter@ptb.de

**Olena Kapran**

## Properties of SFS Josephson junctions with strong ferromagnetic Ni barrier
*O. Kapran, A. Iovan, T. Golod, V. Krasnov*

In this study we investigate Josephson junctions with ferromagnetic Ni as a barrier. The aim of the study is to characterize the junctions from point of view Josephson coupling. We study the thickness dependence of the critical current into a wide range of thicknesses of the ferromagnetic barrier(0-35nm). We see Josephson coupling at high thicknesses of 35nm which open the possibility of building and controlling device based on strong ferromagnetic Josephson junctions. A threshold thickness has been observed at around 4nm where Fraunhofer modulation of the critical field disappears. We attribute the threshold behavior to the physical continuity of ferromagnetic layer. We fabricate and study junctions at different nanoscaled geometries and we investigate the magnetic properties of the ferromagnet based on the electrical properties of the Josephson Junction.

The layers were deposited by physical sputtering and the variation of the thickness of the barrier has achieved by angled deposition. The ferromagnetic barrier is sandwiched between two superconductor layers of the niobium, each 200nm thickness. The junctions are fabricated by FIB (Focused Ion Beam) with geometries in the hundreds of nanometers.

The system will allow the future study of Josephson junctions based on strong ferromagnet spin valves. The spin valve Josephson junction would allow the study of triplet superconductivity and the integration of a wide range of spintronic devices into Josephson junction.

**Olena Kapran**
Department of Physics,
Stockholm University,
AlbaNova University Center,
SE-106 91 Stockholm, Sweden
E-mail: olena.kapran@fysik.su.se

## Supercurrent-induced long-range triplet correlations and controllable Josephson effect in superconductor/ferromagnet hybrids with extrinsic SOC

_A. A. Mazanik[1,2], I. V. Bobkova[3,1,4]_
[1]Moscow Institute of Physics and Technology, Dolgoprudny, 141700 Russia
[2]BLTP, Joint Institute for Nuclear Research, Dubna, Moscow Region, 141980, Russia
[3]Institute of Solid State Physics, Chernogolovka, Moscow region, 142432 Russia
[4]National Research University Higher School of Economics, Moscow, 101000 Russia

It has been shown in Refs. [1,2] that the combination of the condensate motion, the exchange field and the Rashba spin-orbit coupling (SOC) induces generation of the long-range triplet superconducting correlations (LRTC), which can be detected via the unusual behavior of Josephson effect and local density of states in superconductor/ferromagnet structures. In this work we predict that such a generation is a more general phenomenon, which takes place not only for intrinsic SOC, like Rashba SOC, but also for extrinsic impurity-induced SOC. The structure of the supercurrent-induced correlations is studied both for S/F bilayers and S/F/S Josephson junctions with extrinsic SOC in superconductor. We demonstrate that in S/F/S junctions, where the Josephson coupling is realized via the supercurrent-induced correlations, the ground state phase can be switched between 0 and $\pi$. The switching is controlled by relative directions of the condensate momentum in superconducting leads, thus realizing a new physical principle of the 0 $-\pi$ shifter.

Andrei A. Mazanik
PhD student, Moscow Institute of Physics and Technology,
Dolgoprudny, 141700 Russia
Phone: +79199627354
E-mail: mazanandrey@gmail.com
Web: https://www.researchgate.net/profile/Andrew-Mazanik

## Topological features of quantum magnetotransport in $Bi_{1-x}Sb_x$ (0 ≤ x ≤ 0.2) bicrystals

*Fiodor Muntyanu [1], Vitalie Chistol [2], Elena Condrea[1] and Anatolie Sidorenko[1,2]*
[1] *Institute of Electronic Engineering and Nanotechnologies, Chisinau, 2028, Moldova*
[2]*Technical University of Moldova, Chisinau, 2004 Republic of Moldova*

Unusual topological features related to the interface Dirac electrons [1, 2] have been revealed: the longitudinal Hall quasi-plateaus, along with minima in magnetoresistance; the manifestation of Umkehr effect, non- allowed by the crystal symmetry; two new harmonics of quantum transport from interface layers, which characterizes larger than cross-sectional areas of the FS of crystallites; the magnetoresistance peculiarities, indicating both the occurrence of a small group of the infinitely moving electrons and the electronic phase transitions of the semiconductor–semimetal type in magnetic field. A high-field behaviour of $\alpha_{ii}(B)$ has been identified (it linearly increases in magnetic field without saturation, the sign changes from negative to positive, the nontrivial π-Berry phase is observed, etc.) in CIs layers, specifying the signature of 3D topological semimetal at 3D Dirac point forming (x ~ 0.04). In addition, it has been found that the bicrystals of $Bi_{1-x}Sb_x$ (0.07 ≤ *x* ≤ 0.2) alloys exhibit peculiarities typical of 3D TI: $\alpha_{ii}(B)$ undergoes saturation in magnetic field or smoothly increase, the Landau level index *n* in all CIs layers linearly depend on $1/B_n$ and extrapolated to –0.5 if $1/B_n \rightarrow 0$.

**Prof. dr. Fiodor M. Muntyanu**
Institute of Electronic Engineering and Nanotechnologies
2028 Chisinau, Republic of Moldova
Phone: +079454853
E-mail: muntean_teodor@yahoo.com

# Resonant properties of a SQUID consisting of two Phi-0 junctions

*I. Rahmonov[1,2], Yu. M. Shukrinov[1,3]*
[1]*BLTP, Joint Institute for Nuclear Research, Dubna, Moscow Region, Russia*
[2]*Umarov Physical Technical Institute, TAS, Dushanbe, Tajikistan*
[3]*Dubna State University, Dubna, Moscow Region, Russia*

The phase dynamics of a SQUID consisting of two Phi-0 junctions has been investigated [1,2]. The current-phase relation of such junction has a phase shift proportional to the magnetic moment, perpendicular to the gradient of the asymmetric spin-orbital potential. The phase shift allows one to manipulate the internal magnetic moment using the Josephson current. The reverse phenomenon leads to the appearance of the DC component in the superconducting current [3-5]. Based on the numerical solution of the equations of the resistive model for Josephson junctions and the Landau - Lifshitz - Gilbert equations for the magnetic moment of ferromagnetic layer, the current-voltage characteristic and time dependences of the voltage and components of the magnetic moment are calculated. The double resonance in this system is observed, i.e., the simultaneous realization of ferromagnetic resonance and excitation of the SQUID eigenmode. The double resonance in the considered SQUID makes possible the detailed investigation of the resonance properties of Phi-0 junction.

## Acknowledgements

The work was partially funded by the RFBR, according to the project 20-37-70056. Numerical calculations have been done within the framework of the project 18-71-10095 of the Russian National Fund.

**Dr. Ilhom Rahmonov**
Bogoliubov Laboratory of Theoretical Physics,
Joint Institute for Nuclear Research,
Dubna, Moscow Region, 141980, Russia
Phone: +7 4962163734 (off), +7 9057544760 (mobile1)
E-mail: rahmonov@theor.jinr.ru

**Olesya Yu. Severyukhina**

# Modeling of superconducting spin valve magnetic properties

*O. Yu. Severyukhina[1,2], A.Yu., Fedotov [1,2], A.Yu. Salamatina [2], A.V. Vakhrushev[1,2], A.S. Sidorenko[3]*
*[1]Udmurt Federal Research Center of the Ural Branch of the Russian Academy of Sciences, Izhevsk, Russia*
*[2]Kalashnikov Izhevsk State Technical University, Izhevsk, Russia*
*[3]Ghitu Institute of Electronic Engineering and Nanotechnology, Chisinau, Moldova*

Physical and mathematical models have been built for calculating the magnetic properties of a superconducting spin valve [1]. In the report two algorithms for modeling spin dynamics are presented: without lattice vibrations and with allowance for lattice vibrations [2]. Mathematical modeling was carried out in the LAMMPS software package using modern empirical many-particle potentials EAM (embedded atom method).

Methods for determining the magnetic properties of nanostructured materials are presented. The general form for describing the total energy of magnetic systems can be written in the sum of the Zeeman and exchange interactions, the magnetic anisotropy, the Dzyaloshinsky-Moriya interactions, magnetoelectric and dipole.

The properties superconducting spin valve, such as magnetization, spin temperature, and exchange interaction energy, have been calculated.

## Acknowledgements

The study was financially supported by the Russian Science Foundation Grant (RSF) Nr. 20-62-47009.

**Dr. Olesya Yu. Severyukhina**
Baramzinoy 34,
426067 Izhevsk, Russia
Phone: +7 3412-21-45-83
Fax: + 7 3412- 50-79-59
E-mail: lesienok@mail.ru
Web: http://iam.udman.ru/en

# Controllable two- *vs* three-state magnetization switching in single-layer epitaxial $Pd_{1-x}Fe_x$ films and $Pd_{0.92}Fe_{0.08}$/Ag/$Pd_{0.94}Fe_{0.04}$ heterostructure

*I. V. Yanilkin[1], A. I. Gumarov[1], G. F. Gizzatullina[1], R. V. Yusupov[1], L. R. Tagirov[2,3]*

[1]*Institute of Physics, Kazan Federal University, Kazan, Russia*
[2]*Zavoisky Physical-Technical Institute, FRC Kazan Scientific Center of RAS, Kazan, Russia*
[3]*Tatarstan Academy of Sciences, Institute of Applied Research, Kazan, Russia*

The generation of the long-range triplet component of the superconducting pairing at noncollinear orientations of magnetizations in ferromagnetic layered systems is extensively studied in magnetic Josephson junctions (MJJ) [1,2] and superconductive spin valves (SSV) [3]. Palladium-iron ($Pd_{1-x}Fe_x$) alloy with $x < 0.10$ has strong practical interest for such MJJ and SSV structures as a material for weak ferromagnetic layers with tunable magnetic properties [4]. Therefore, it is useful to investigate the switching properties of $Pd_{1-x}Fe_x$ films and heterostructures with the aim to create controllable non-collinear magnetic states.

Single epitaxial film $Pd_{0.92}Fe_{0.08}$(20 nm) (hereinafter PdFe1) and epitaxial heterostructure $Pd_{0.92}Fe_{0.08}$(20 nm)/Ag(20 nm)/$Pd_{0.94}Fe_{0.04}$(20 nm) (hereinafter PdFe1/Ag/PdFe2) were grown using molecular beam deposition technique [5]. Detailed measurements of the magnetoresistance have shown that the PdFe1 epitaxial film, being an easy-plane ferromagnet with a pronounced in-plane anisotropy, undergoes magnetization switching between two (with collinear magnetization directions) or three (including orthogonal to the two directions noted above) single-domain states, depending on the direction of the applied magnetic field. This unconventional magnetization switching is a consequence of a combination of the fourfold magnetocrystalline anisotropy and additional uniaxial magnetic anisotropy in the film plane.

The use of two magnetic layers PdFe1 and PdFe2 with different coercive fields, separated by a nonmagnetic spacer, makes it possible to obtain parallel, orthogonal, and antiparallel configurations of magnetic moments. It has been experimentally demonstrated that the PdFe1/Ag/PdFe2 heterostructure can be switched between three stable magnetic configurations in the film plane by rotating the magnetic moment of the soft magnetic layer with respect to the hard magnetic layer. Such a structure can serve as a magnetic actuator for switching an MJJ from the singlet conducting mode to the triplet conducting mode and vice versa.

**Dr. Igor V. Yanilkin**
Institute of Physics, Kazan Federal University,
Kremlevskaya18
420008 Kazan, Russia
Phone: +7 843 2337763
E-mail: yanilkin-igor@yandex.ru
Web: https://kpfu.ru/igor.yanilkin?p_lang=2

# Ultrafast Optical and Magnetooptical Manifestations of Nanoscale Magnetic Inhomogeneities in Epitaxial $Pd_{1-x}Fe_x$ Thin Films

_R. V. Yusupov[1], A. V. Petrov[1], S. I. Nikitin[1], I. V. Yanilkin[1], A. I. Gumarov[1], L. R. Tagirov[2,3]_
[1]_Institute of Physics, Kazan Federal University, Kazan, Russia_
[2]_Zavoisky Physical-Technical Institute, FRC Kazan Scientific Center of RAS, Kazan, Russia_
[3]_Tatarstan Academy of Sciences, Institute of Applied Research, Kazan, Russia_

Josephson superconductor-ferromagnet hybrids are the key elements of superconducting spintronics [1-3]. Palladium-iron ($Pd_{1-x}Fe_x$) alloy with $x < 0.10$ is a practical material for the phase inverting ferromagnetic layers in these hybrids [2,3], and magnetic homogeneity of the alloy is one of the critical requirements providing low damping of the superconducting pairing function and high critical current. In particular, film composition ($x$) and conditions of synthesis should be defined that ensure its magnetic homogeneity. Conventional neutron scattering methods can hardly be used to probe magnetic inhomogeneities in one-two dozens of nanometer thick films. Stationary methods like DC magnetometry, magnetooptical Kerr and ferromagnetic resonance techniques reflect the integral properties of the ferromagnetic fraction and are not sensitive to nanometer-scale inhomogeneities.

Here we report on a study of a series of $Pd_xFe_{1-x}$ films with $x$ = 0, 0.03, 0.06 and 0.08 prepared by molecular beam epitaxy, each 20 nm thick. The ultrafast laser pump-probe spectroscopies in a 4 – 300 K temperature range [4] reveal the manifestations of the transition to ferromagnetic state both in the reflectivity and the magnetooptical Kerr angle (MOKE) transients. Photoinduced demagnetization below the Curie temperature $T_C$ proceeds in two stages: the subpicosecond stage present at any temperature, and a slower, 10-25 ps, stage reflecting the ferromagnet/paramagnet (FM/PM) type inhomogeneity inherent to the palladium-rich $Pd_{1-x}Fe_x$ alloys. Another signature of the PM-phase occurrence below the $T_C$ is the slow, 0.4-1.0 ns, reflectivity relaxation component, whose magnitude variation with temperature brings to the estimates of the residual PM-phase volume fraction of $\sim$ 30% for $x$ = 0.03 and $\sim$ 15% for $x$ = 0.06 films. The minimal iron content ensuring the magnetic homogeneity of the low-temperature FM-state in epitaxial $Pd_{1-x}Fe_x$ alloy films is about 8 at.%. The mechanism of the nearly-localized $d$-electron ballistic transport between the PM and FM regions is discussed for the slower demagnetization component. This mechanism is significant for a mixed FM/PM system with the nanometer length scale of inhomogeneity.

**Assoc. Prof. Dr. Roman V. Yusupov**
Institute of Physics, Kazan Federal University,
Kremlyovskaya str. 18
420008 Kazan, Russia
Phone: +7 917-3942-192
E-mail: Roman.Yusupov@kpfu.ru
Web: https://kpfu.ru/Roman.Yusupov

# Long-range interaction of magnetic moments in a coupled system of S/F/S Josephson junctions with anomalous ground state phase shift

_G. A. Bobkov[1], A. M. Bobkov[2], I. V. Bobkova[2,1,3]_

[1]_Moscow Institute of Physics and Technology, Dolgoprudny, Russia_
[2]_Institute of Solid State Physics, Chernogolovka, Russia_
[3]_National Research University Higher School of Economics, Moscow, Russia_

A mechanism of a superconductivity-mediated interaction of two magnets in a system of coupled superconductor/ferromagnet/superconductor Josephson junctions (JJs) with spin-orbit interaction is proposed [1]. It is based on the magnetoelectric coupling between the condensate phase difference and the magnetization in the interlayer of the S/F/S JJ, which is realized in the form of the anomalous ground state phase shift. The current-phase relation of the S/F/S takes the form $j = j_c sin(\varphi - \varphi_0)$, where $\varphi_0$ is the anomalous phase shift, which is proportional to the magnetization component perpendicular to the current. The energy of the coupled system and the equilibrium mutual orientation of the ferromagnetic interlayers depend on the external superconducting phase χ between the leads. Thus, it is shown that there is an effective coupling between the magnets and the coupling depends on the superconducting phase. The interaction between the magnets is mediated by the superconducting phase of the middle superconductor, which is a macroscopic quantity and interacts with both magnetizations in the presence of the anomalous phase shift. The coupling is long-range because the length of the middle superconductor is not restricted by the superconducting coherence length. The magnetic configuration can be dynamically manipulated by $\chi$.

**Grigorii A. Bobkov**
Moscow Institute of Physics and Technology,
Dolgoprudny, Russia
E-mail: bobkov.ga@phystech.edu

## Magnetothermopower features in bismuth wires at 80 K

*Elena Condrea[1], Fiodor Muntyanu[1] and Andrzzej Gilewski[3]*
*[1]D.Ghitu Institute of Electronic Engineering and Nanotechnologies, Chisinau, R. Moldova*
*[2]MagNet, 50-421 Wroclaw, Poland*

The presented investigations of the magnetotransport measurements of Bi wires complement the series of recently published experimental results on bismuth nanowires in the deferent experimental configurations of magnetic field respective to the crystal orientation [1,2,3].

Measurements of the transverse magnetothermopower and magnetoresistance in single bismuth wires are performed at 80 K. Appearance of the asymmetry in the magnetothermopower defined as Umkher effect is observed. The investigations is focused on the manifestation of the Umkehr effect in the thermopower under the condition of the Electronic Topological Transition of the Fermi surface at the application of a directed deformation along bisectrix axis.

Study of the magnetotransport properties under uniaxial strain revealed a reorganization of the band structure followed by the changes in the anisotropy of the Fermi surface at high strain values. A moderate applied stress leads to a decrease in the magnitude of the Umkehr effect in the thermopower up to the disappearance of the effect along one of the principal crystallographic axes. The observed behavior of the magnetothermopower and magnetoresistance is explained in the framework of the phenomenological theory of transport phenomena in Bi.

**Key words:** bismuth, magnetoresistance, Seebeck effect, high magnetic field, strain

**Assoc. Prof. Elena Condrea**
Head of Cryogenic laboratory,
D.GHITU Institute of Electronic Engineering and Nanotechnologies,
Chisinau, Republic of Moldova
E-mail: condrea@nanotech.md

# Peroxidase-Like Properties of Fe$_3$O$_4$/PVP Nanocompozite in Hydrogen Peroxide Detection

_**T.D. Gutsul**, E.G. Coscodan, A.A. Sirbu, D.S. Nirca_
_D.Ghitu Institute of Electronic Engineering and Nanotechnologies, Chisinau, Moldova_

Nanozymes are nanomaterials exhibiting intrinsic enzyme-like activity. They have found applications in various fields, such as biosynthesis, environmental protection, and bioelectronic sensing devices [1]. However, artificially designed biomimetic nanomaterials mostly exhibit poor dispersion and low catalytic activity. Therefore, the search for new approaches to designing inexpensive biomimetic materials is inevitable. To this end, a nanocomposite containing magnetite nanoparticles and the poly-*N*-vinylpyrrolidone (PVP) polymer—Fe$_3$O$_4$/PVP—was synthesized by the solvothermal method. The peroxidase activity of the synthesized Fe$_3$O$_4$/PVP nanocomposite as a function of concentration was studied; the colorometric detection of hydrogen peroxide was simulated in accordance with the developed standard [2].

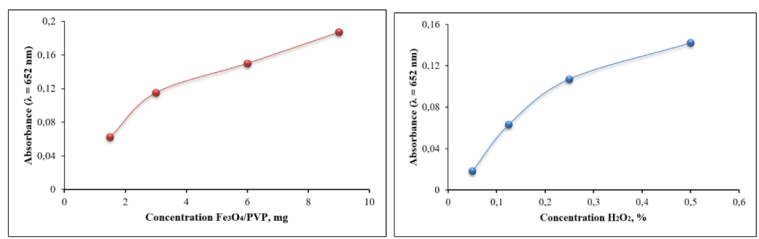

*Fig 1. (A) Colorometric study of the dependence of peroxidase activity on the concentration of the Fe$_3$O$_4$/PVP nanocomposite and (B) dose–response curve for H$_2$O$_2$ Fe$_3$O$_4$/PVP*

Figure 1 shows results of the studies; the observed increase in absorbance as a function of the magnetite nanoparticle concentration suggests that the enzyme-like nanocomposite exhibits the peroxidase activity, while the increase in absorbance as a function of hydrogen peroxide concentration in the presence of the nanocomposite makes it possible to use this catalytic process to detect the presence of H$_2$O$_2$ in various objects, such as rainwater.

**Tatiana Gutsul**
D. Ghitu Institute of Electronic Engineering and Nanotechnologies, Chisinau, Moldova
E-mail: tatiana.g52@mail.ru

# Responsivity and detectivity of $Zn_{0.8}Mg_{0.2}O$/p-Si prepared by spin coating and aerosol deposition method

*V. Morari [1], E. Rusu [1], V. Ursaki [2], I. M. Tiginyanu [2]*
*[1]D. Ghitu Institute of Electronic Engineering and Nanotechnologies, Chisinau, Moldova*
*[2]National Center for Materials Study and Testing, Technical University of Moldova, Chisinau*

ZnMgO solid solutions system presents interest for optoelectronic application due to possibilities to tailor many important physical properties by varying their composition [1]. In this paper, we present data concerning responsivity and detectivity of $Zn_{0.8}Mg_{0.2}O$/p-Si photodetectors based on thin films prepared by spin-coating and aerosol deposition methods from 0.35 M aqueous solution using Zn and Mg acetates. The morphological and chemical composition of films has been investigated in details by scanning electron microscopy (SEM) and Energy Dispersive X-ray analysis (EDX). The photoelectrical parameters of detectors (Fig. 1) have been deduced from current-voltage characteristics measured in the dark and under UV illumination.

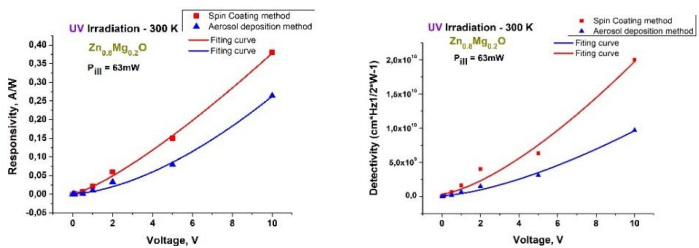

*Figure 1. The responsivity and detectivity under different bias voltage of ZnMgO/p-Si thin films*

The photocurrent was generated by a xenon DKSS-150 lamp using an optical filter (УФС5-300 nm). The responsivity (R) and detectivity (D*) under UV illumination of 63 mW/cm$^2$ and 10 V direct bias were found to be of R = 0.38 A/W, D* = $2\times10^{10}$ cm×Hz$^{1/2}$W$^1$ for films deposited by spin coating; and R = 0.264 A/W, D$^*$ = $9.7\times10^9$ cm×Hz$^{1/2}$W$^{-1}$ for films prepared by aerosol deposition.

**Acknowledgements**
This work was supported financially by the National Agency for Research and Development, Republic of Moldova, through grant No. *20.80009.5007.02.*

**Vadim Morari**
D. Ghitu Institute of Electronic Engineering and Nanotechnologies
Chisinau, Moldova
E-mail: vadimmorari2018@gmail.com

## Modeling of cluster ion beams implantation into a metal substrate

*S.V. Suvorov [1], A.V. Vakhrushev[1,2] A.S. Sidorenko[3]*
*[1]Udmurt Federal Research Center of the Ural Branch of the Russian Academy of Sciences, Izhevsk, Russia*
*[2]Kalashnikov Izhevsk State Technical University, Izhevsk, Russia*

Modern methods for the synthesis of nanocomposites do not allow simultaneous control of their chemical, morphological, and microstructural properties. By H. Khan, T. Reisinger and their colleagues [1] a new method was developed for creating fully adaptable nanocomposites by combining cluster ion beams and thin-film deposition technologies for the manufacture of cluster nanocomposites under well-defined conditions. These processes are very complex; therefore, the application of mathematical modeling methods to their analysis is highly relevant.

In this work, on the basis of mathematical modeling in nanosystems [2], a technique has been developed for modeling the implantation of cluster ion beams into a metal substrate. The results of numerical calculations at different substrate temperatures and velocities of cluster ion beams are presented.

**Dr. Stepan V. Suvorov**
Baramzinoy 34,
426067 Izhevsk, Russia
Phone: +7 3412-21-45-83
Fax: + 7 3412- 50-79-59
E-mail: ssv.82@mail.ru
Web: http://iam.udman.ru/en

# Colorimetric biosensor based on ZnO / ZnFe$_2$O$_4$ heterostructures

**A. A. Sirbu\*, D. S. Nirca, T. D. Gutul, V. M. Fedorov and A. S. Sidorenko**

*Institute of Electronic Engineering and Nanotechnologies 'D. Ghitu', 3/3 Academie, MD-2028 Chisinau, Republic of Moldova*

Currently, nanomaterials based on inorganic nanoparticles are widely used in biology, medicine, biotechnology, food industry, and agriculture due to their unique properties to exhibit enzyme-like activity. The relevance of this report is determined by the fact that diagnostic devices for the specific detection of biological and chemical targets at low concentrations in a reliable, convenient and inexpensive manner Of great interest is the design of sensitive sensors based on nanozymes, i.e., enzymatic nanomaterials, such as ZnO, Fe$_3$O$_4$, and ZnFe$_2$O$_4$. These materials have high sensory properties and can catalyze the reaction of the peroxidase substrate to change the color of the 3,3',3,5'-tetramethylbenzidine (TMB) dye [1]. In our study, nanoparticles ZnO of size 30-40 nm were by the sol – gel method and with the solvatothermal method obtained ZnFe$_2$O$_4$ with a dimension of 5–6 nm with poly(N-vinylpyrrolidone) (PVP) as a stabilizer. The obtained nanoparticles were characterized by X-ray powder diffraction (XRD) and scanning electron microscopy (SEM). ZnO gives us very good photocatalytic properties, which affect the catalytic reaction occurring in the presence of ZnFe$_2$O$_4$ nanocomposite, these photocatalytic properties will contribute to accelerate the reaction after creating the heterostructure underlying this work. These photocatalytic properties contribute to the reaction acceleration after the creation of the heterostructure underlying this work. Based on these nanoparticles, a heterostructure was prepared by casting and used for glucose detection. The study was conducted by colorimetric method at a wavelength of λ = 652 nm to visualize the redox reaction occurring in a model system containing solutions of glucose oxidase (GOx), dye TMB and glucose [2]. The study showed the possibility of glucose detection in the concentration range of 1 - 11 mM using the ZnO/ZnFe$_2$O$_4$ heterostructure.

**Andrei Sirbu**
Institute of Electronic Engineering and Nanotechnologies „D. Ghitu"
3/3 Academie, MD-2028 Chisinau, Republic of Moldova
E-mail: andrei.sirbu721@gmail.com

# Millimeter wave to terahertz compact and low-loss superconducting plasmonic waveguides for cryogenic integrated nano-photonics

Samane Kalhor [1,2], Majid Ghanaatshoar [2], Hannah J. Joyce [3], David A. Ritchie [4], Kazuo Kadowaki [5] and Kaveh Delfanazari [1,3,4*]

[1] James Watt School of Engineering, University of Glasgow, Glasgow G12 8QQ, UK

[2] Laser and Plasma Research Institute, Shahid Beheshti University, G.C., Evin, 1983969411, Tehran, Iran

[3] Electrical Engineering Division, University of Cambridge, Cambridge CB3 0FA, UK

[4] Department of Physics, Cavendish Laboratory, University of Cambridge, Cambridge CB3 0FA, UK

[5] Division of Materials Science, Faculty of Pure & Applied Sciences, University of Tsukuba 1-1-1, Tennodai, Tsukuba, Ibaraki 305-8573, Japan

[*] Corresponding author: kaveh.delfanazari@glasgow.ac.uk

Plasmonic, as a rapidly growing research field, provides new pathways to guide and modulate highly confined light in the microwave to the optical range of frequencies. We demonstrate a plasmonic slot waveguide, at the nanometer scale, based on high transition temperature (Tc) superconductor $Bi_2Sr_2CaCu_2O_{8+\delta}$ (BSCCO), to facilitates the manifestation of the chip-scale millimeter waves (mm- waves) to terahertz (THz) integrated circuitry operating at cryogenic temperatures. We investigate the effect of geometrical parameters on the modal characteristics of the BSCCO plasmonic slot waveguide between 100 GHz and 500 GHz. In addition, we investigate the thermal sensing of the modal characteristics of the nanoscale superconducting slot waveguide and show that at a lower frequency, the fundamental mode of the waveguide has a larger propagation length, a lower effective refractive index, and a strongly localized modal energy. Moreover, we find that our device offers a larger SPP propagation length and higher field confinement than the gold plasmonic waveguides at broad temperature ranges below BSCCO's Tc. The proposed device can open up a new route towards realizing cryogenic low-loss photonic integrated circuitry at the nanoscale.

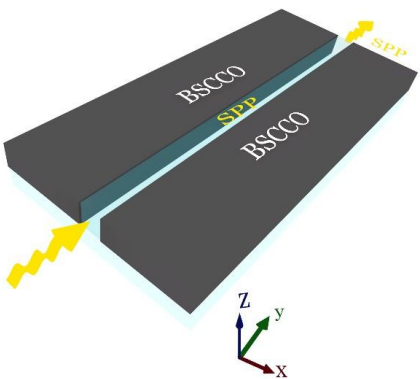

Figure 1: 3D Schematic diagram of the proposed BSCCO based plasmonic slot waveguide

**Dr. Kaveh Delfanazari**
Electronics and Nanoscale Engineering (ENE) Division, School of Engineering, University of Glasgow, Glasgow, G12 8QQ UK
E-mail: Kaveh.Delfanazari@glasgow.ac.uk

Kaveh Delfanazari

# Voltage-controlled Polarization and Cavity modes in Integrated Superconducting Coherent Terahertz Emitters

**Yusheng Xiong**[1], **Takanari Kashiwagi**[2], **Richard A Klemm**[3], **Kazuo Kadowaki**[2], **Kaveh Delfanazari**[1*]

[1]*James Watt School of Engineering, University of Glasgow, Glasgow G12 8QQ, UK*

[2]*Division of Materials Science, Faculty of Pure & Applied Sciences, University of Tsukuba 1-1-1, Tennodai, Tsukuba, Ibaraki 305-8573, Japan*

[3]*Department of Physics, University of Central Florida, Orlando, FL 32816, USA*

[*]Corresponding author: kaveh.delfanazari@glasgow.ac.uk

Solid-state, compact and integrated terahertz (THz) devices based on high-$T_C$ superconducting $Bi_2Sr_2CaCu_2O_{8+\delta}$ (BSCCO) can coherently and continuously radiate electromagnetic waves with frequencies tunable between 100 GHz and 11 THz. The high power *cw* THz wave can be observed by the application of an applied voltage of as small as $V_{dc}$ (V) <1.5 through the device plane. The frequency tunability of such chip-scale quantum devices spans the entire THz gap. Here, we report on a novel approach towards control of THz waves in superconducting THz emitters with pentagonal cavities. We perform numerical simulations/analytical calculations and study their cavity resonances and linear to circular polarization. We engineer the radiation of the intense and coherent THz waves in pentagonal emitters in various ways and compare them with experimental results. We also control the cavity modes and polarization by fixing the bias feed point and changing the device geometry.

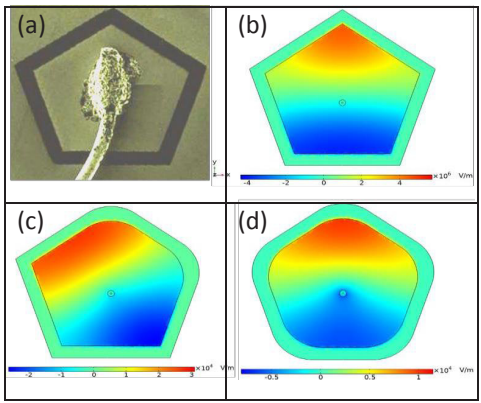

*Fig. 1. (a) A false color scanning ion microscope image of the superconducting THz emitter with regular pentagonal cavity. The electrical bias feed line is located in the middle of the device. Sketch of the pentagonal THz devices together with the electric field $E_z$ map at f = 0.46 THz., for (b) regular pentagonal geometry similar to that fabricated THz device shown in (a), (c) for regular pentagonal cavity with two rounded edges, and (d) for regular pentagonal cavity with all rounded edges. In (b-d) a single-feed point is chosen to be in the center.*

**Dr. Kaveh Delfanazari**

Electronics and Nanoscale Engineering (ENE) Division, School of Engineering, University of Glasgow, Glasgow, G12 8QQ UK

E-mail: Kaveh.Delfanazari@glasgow.ac.uk

# The use of low-intensity current without external power supplies for the treatment of patients with burns

_R. Chornopyshchuk[1], V. Nagaichuk[1], O. Nazarchuk[2], V. Nagaichuk[3], S. Sidorenko[4], L. Sidorenko[5]_

[1]National Pirogov Memorial Medical University, Department of General Surgery, Vinnytsya, Ukraine

[2]National Pirogov Memorial Medical University, Department of Microbiology, Vinnytsya, Ukraine

[3]National Pirogov Memorial Medical University, Department of Surgical Stomatology, Vinnytsya, Ukraine

[4]Rehazentrum Valens, Valens, Switzerland

[5]State Medical and Pharmaceutical University N. Testemitanu, Department of Molecular Biology and Human Genetics, Chisinau, Moldova

Recently, along with traditional medicines for the treatment of burns, the possibility of using physical methods that can affect various stages in the pathogenesis of the wound process is of great interest [1]. The use of low-intensity currents without external power supplies, which are generated directly by biological objects, deserves special attention [2]. Despite the fact that experimental research in this direction began only in the middle of the twentieth century, convincing positive results were obtained for the effectiveness of their use, which contributed to its rapid introduction to practical medicine in various fields [3]. Electroreflexology and electrophoresis of drugs were the forms implemented. Further analysis of the results of clinical application of electroreflexology allowed to identify a new direction in medicine - biogalvanization, which on the basis of cutaneous and visceral interactions allows to identify, correct and promote normalization of autonomic functional and energy balance, while ensuring bioenergetic homeostasis. A significant contribution to the study of this area was made by prof. Makats V.G. et al., who were particularly interested in application of this method in combustiology, especially taking into account the integral function of the skin [4]. It was found that direct galvanic current is an active biological stimulator, which is realized with the participation of humoral, neuroreflectory and immune mechanisms [5]. In addition, this current is able to promote the penetration of drugs into the body. Such directional electrophoresis allows to provide high concentrations of active substances in the wound, enhancing their pharmacological effect. Moreover, the ability of low-intensity currents to inhibit the growth and reproduction of microorganisms, as well as increase their sensitivity to antimicrobial agents, has been established [6].

The use of low-intensity currents without external power supplies in combination with existing traditional treatment of patients with wounds, as wound dressings, remains promising. The developed technology of activation of lyophilized pig skin xenografts by biogalvanic current for closing burn wounds at early surgical necrectomies allowed to improve efficiency of their application with a possibility of longer presence on a wound [7]. So far, positive results of _in vitro_ experimental studies have been obtained to enhance the antimicrobial properties of synthetic composite modified materials based on poly(2-hydroxyethyl methacrylate) with the addition of drugs [8].

Thus, the use of low-intensity currents without external power supplies in medicine and in the treatment of patients with burns in particular, has undeniable effectiveness and requires wider implementation in clinical practice. The lack of information on a number of mechanisms of action of such currents on biological objects unequivocally indicates the need for further research in this direction.

**Dr. Roman Chornopyshchuk**
National Pirogov Memorial Medical University
Department of General Surgery
Pirogov str. 56
21028 Vinnytsya, Ukraine
Phone: +38 097-2128-963
Fax: +38 0432-67-01-91
E-mail: r.chornopyshchuk@gmail.com
Web: https://www.vnmu.edu.ua/en/department-of-general-surgery

# Emergent Interfacial Superconductivity between Twisted Cuprate Superconductors

*S. Y. F. Zhao[1], N. Poccia[1,2], X. Cui[1], P. A. Volkov[3], H. Yoo[1], R. Engelke[1], Y. Ronen[1], R. Zhong[4], G. Gu[4], S. Plugge[5], T. Tummuru[5], M. Franz[5], J. H. Pixley[3], P. Kim[1]*

[1]*Department of Physics, Harvard University, Cambridge, MA 02138, USA*

[2]*Institute for Metallic Materials, IFW Dresden, 01069 Dresden, Germany*

[3]*Department of Physics and Astronomy, Center for Materials Theory, Rutgers University, Piscataway, NJ 08854, USA*

[4]*Department of Condensed Matter Physics and Materials Science, Brookhaven National Laboratory, Upton, NY 11973, USA*

[5]*Department of Physics and Astronomy & Stewart Blusson Quantum Matter Institute, University of British Columbia, Vancouver, BC, Canada*

Twisted interfaces between stacked van der Waals cuprate crystals enable tunable Josephson coupling between in-plane anisotropic superconducting order parameters. Employing a novel cryogenic assembly technique, we fabricate Josephson junctions with an atomically sharp twisted interface between $Bi_2Sr_2CaCu_2O_{8+x}$ crystals. The Josephson critical current density sensitively depends on the twist angle, reaching the maximum value comparable to that of the intrinsic junctions at small twisting angles, and is suppressed by almost 2 orders of magnitude yet remains finite close to 45 degree twist angle. Through the observation of fractional Shapiro steps and the analysis of Fraunhofer patterns we show that the remaining superconducting coherence near 45 degree is due to the co-tunneling of Cooper pairs, a necessary ingredient for high-temperature topological superconductivity.

**Prof. Philip Kim**
Department of Physics Harvard University
11 Oxford Street, Cambridge MA 02138, USA
Phone: +1 617-496-0714
Fax: +1 617-495-0416
E-mail: pkim@physics.harvard.edu
Web: https://www.ifw-dresden.de/about-us/people/prof-dr-vladimir-fomin

**Leonid Konopko**

# Quantum oscillations in microwires of topological insulator contacted with superconducting leads

*L. Konopko[1], A. Nikolaeva[1], T. Huber[2], K. Rogacki[3]*

[1]*D.Ghitu Institute of Electronic Engineering and Nanotechnology, Chisinau, Moldova*
[2]*Department of Chemistry, Howard University, DC 20059, Washington, U.S.A.*
[3]*Institute of Low Temperature and Structure Research, PAS, Wroclaw 50950, Poland*

Recent efforts to detect and manipulate Majorana fermions in solid state devices have employed topological insulator (TI) nanowires proximity coupled to superconducting leads (SC). We studied the transverse magnetoresistance of $Bi_2Te_3$ and $Bi_{0.83}Sb_{0.17}$ TI microwires contacted with superconducting $In_2Bi$ leads [1]. The equidistant in transverse magnetic field (up to 1 T) magnetoresistance (MR) oscillations at the TI/SC interface have been observed at various temperatures (4.2 − 1.5 K) both in $Bi_2Te_3$ and in $Bi_{0.83}Sb_{0.17}$ samples. The oscillations almost disappear when the measurement temperature reaches the superconducting transition temperature in $In_2Bi$. The amplitude of the MR oscillations also decreases with increasing magnetic field; in magnetic fields of $B > 0.6$ T at $T = 1.5$ K, oscillations are not visible. Possibly, the observed oscillations can be the Aharonov-Bohm oscillations of the magnetic flux quantization. In this case, a closed trajectory is formed at the edge states of the TI/SC interface. The oscillation period $\Delta B=(h/e)/S$ were $h/e$ is flux quantum, $S$ is the cross-sections area of closed trajectory; then, taking into account the oscillation periods, for $Bi_2Te_3$ microwire the effective trajectory diameter should about 500 nm, while for $Bi_{0.83}Sb_{0.17}$ microwire − about 300 nm. In both cases, these diameters are much smaller than the corresponding microwire diameters. Different assumptions about the nature of the observed effect are discussed.

This work was supported by the Moldova State Project # 20.80009.50007.02, NSF through STC CIQM 1231319, the Boeing Company and the Keck Foundation.

**Dr. Leonid Konopko**
Electronics of Low Dimensional Structures Lab.,
Institute of Electronic Engineering and
Nanotechnologies
Academiei str. 3/3, MD-2028,
Chisinau, Republic of Moldova
Phone: +37322 737072
E-mail: l.konopko@nanotech.md
Web: https://nanotech.md

**Maria Lupu**

## Single-Layer Perceptron: the basic principles of construction and functioning

_M.C. Lupu[1]_
_[1]Ghitu Institute of Electronic Engineering and Nanotechnologies, Chisinau, Republic of Moldova_

In machine learning, the perceptron is an algorithm for supervised learning of binary classifiers. A binary classifier is a function which can decide whether or not an input, represented by a vector of numbers, belongs to some specific class. It is a type of linear classifier, a classification algorithm that makes its predictions based on a linear predictor function combining a set of weights with the feature vector.

The perceptron algorithm was invented in 1958 at the Cornell Aeronautical Laboratory by Frank Rosenblatt, funded by the United States Office of Naval Research.

The perceptron was intended to be a machine, rather than a program, and while its first implementation was in software for the IBM 704, it was subsequently implemented in custom-built hardware as the "Mark 1". This machine was designed for image recognition: it had an array of 400 photocells, randomly connected to the "neurons". Weights were encoded in potentiometers, and weight updates during learning were performed by electric motors.

Single-layer perceptrons are only capable of learning linearly separable patterns. For a classification task with some step activation function, a single node will have a single line dividing the data points forming the patterns. More nodes can create more dividing lines, but those lines must somehow be combined to form more complex classifications. A second layer of perceptrons, or even linear nodes, are sufficient to solve a lot of otherwise non-separable problems.

The simplest kind of neural network is a single-layer perceptron network, which consists of a single layer of output nodes; the inputs are fed directly to the outputs via a series of weights. The sum of the products of the weights and the inputs is calculated in each node, and if the value is above some threshold (typically 0) the neuron fires and takes the activated value (typically 1); otherwise it takes the deactivated value (typically -1). Neurons with this kind of activation function are also called artificial neurons or linear threshold units. In the literature the term perceptron often refers to networks consisting of just one of these units. A similar neuron was described by Warren McCulloch and Walter Pitts in the 1940s.

Having considered the basic principles of the structure and functioning of a single-layer perceptron, it is possible to draw conclusions about the possibility of its application in various fields and improvement of performance in order to create models that are able to perform the assigned tasks as accurately as possible and to learn as quickly as possible.

**Maria Lupu**
PhD student, D. Ghitu Institute of Electronic Engineering and Nanotechnologies
Chisinau,Republic of Moldova
E-mail: simplex_prof_100@mail.ru

# Influența radiației ultraviolete bactericide asupra componentelor structurale ale genomului virusului SARS – COV – 2

*Iu. N. Nica, L.B. Pogorelischii, S.N. Zavrajny, A.S. Sidorenko*

*Institutul de Inginerie Electronică şi Nanotehnologii D. Ghitu, str. Academiei 3/3, MD-2028, Chişinău, Moldova*

Pandemia de COVID19 care terorizează lumea cu o agresivitate teribilă solicită găsirea urgentă a instrumentelor care ar inactiva rapid viruşii din mediul ambiant pentru a reduce şansa de infecţie prin aerosoli şi transmiterea prin contact.  Pentru inactivarea viruşilor SARS–CoV–2 am utilizat LED-uri cu emisia maximală pe lungimea de undă 255 ± 5nm. Aplicând metoda reacţiilor în lanţ a polimerazei în timp real cu transcripţie inversă (RT qPCR) am cercetat influenţa radiaţiei UVC asupra unităţilor structurale ale genomului SARS-CoV-2: genele N şi E (specifice coronavirusului SARS-CoV-2). Pentru iradierea materialului care conţine COVID-19, a fost elaborată şi fabricată o instalaţie care include:

1. O masă pentru plasarea de diapozitive sau cutii Petri cu probe, un suport pentru o matrice LED şi un obturator electromagnetic, asamblate într-un singur bloc.

2. Unitatea de control, care include un temporizator pentru controlul obturatorului, un stabilizator de curent prin matricea LED. Alimentarea dispozitivului de la ~ 220V AC şi stabilizarea modurilor de funcţionare a circuitelor electrice interne este asigurată de un adaptor AC / DC 220V / 30V 2A.

Fiecare experiment s–a petrecut în felul următor:

a)selectarea materialului care conţinea viruşi SARS-CoV-2 şi prepararea probelor împărţind materialul selectat în părţi egale şi fiecare parte se depune într – un strat subţire pe sticla special pregătită.

b) iradierea probelor cu diferite doze energetice de radiaţie UVC (una din probe nu se iradiază, pentru a putea aprecia concentraţia iniţială a viruşilor).

c) efectuarea procedurilor de amlificare numerică a componentelor structurale ale viruşilor prin metoda RT qPCR.

Prezentăm rezultatele procedurilor RT qPCR cu mostre de SARS–CoV–2 tratate cu radiaţii UVC cu diferite doze energetice.

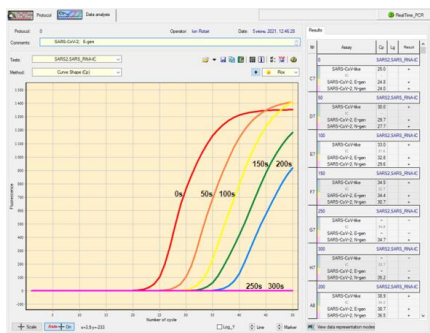 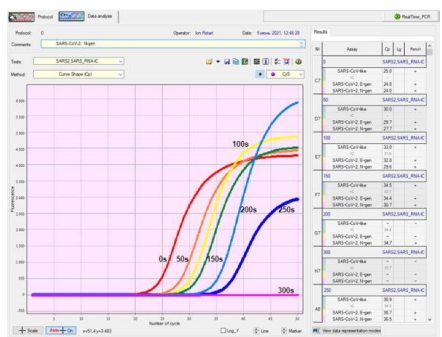

*Fig. 1 Gena E*              *Fig. 2 Gena N*

Pe Figurile 1 si 2 este prezentată dependenţa fluorescenţei de numărul ciclurilor de amplificare pentru mostrele iradiate cu diferite doze energetice (0, 200, 400, 600, 800, 1000, 1200 mJ/cm$^2$). Axa verticală prezintă intensitatea fluorescenţei (numărul de fluorofori ataşaţi componentelor structurale ale viruşilor), iar axa orizontală – numărul ciclurilor de amplificare.

Observăm, că odată cu mărirea dozei energetice cu care sunt tratate mostrele, crește numărul ciclurilor după care începe să se manifeste amplificarea numărului genelor. La doza corespunzătoare iradierii 1,0 J/cm$^2$ și mai mult, gena E este complet anihilată (curbele corespunzătoare dozelor 1 și 1,2 J/cm$^2$ sunt amplasate pe orizontală). Pentru gena N amplificarea nu se mai manifestă începând cu doza energetică de 1,2 J/cm2. Datele obținute în timpul experimentelor au arătat că iradierea probelor cu lumină ultravioletă la 255 nm cu o densitate de energie 280 mJ/cm$^2$ reduce conținutul genei E de 100 ori.

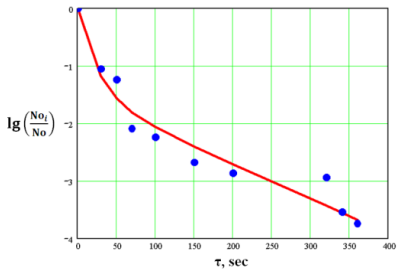

Pentru a reduce cantitatea acestei gene cu un factor de ~ 1000, este necesară o expunere cu 1,2 J/cm$^2$. Iar expunerea la radiații cu ~ 1,6 J/cm$^2$ va reduce cantitatea genei E de 10 000 ori. (Fig 3).

Studiul a fost efectuat în cadrul proiectelor: 20.70086.07/COV(70105) ,,Elaborarea dispozitivelor pentru dezinfectarea vestimentației și echipamentului personalului care intră în contact cu purtătorii de SARS-CoV-2 (TEHMED COVID)" din Programul UE Orizont 2020 (**CG-ENI/2017/ 386-980 CE** „Suportul Financiar pentru participarea Republicii Moldova în Programul Cadru al Uniunii Europene de cercetare și inovare ORIZONT 2020)", și 2. 65/22.10.19 A „Nanostructuri și nanomateriale funcționale pentru industrie și agricultură" din Programul de Stat.

**Dr. Iurie N. Nica**
Institutul de Inginerie Electronică și Nanotehnologii D. Ghitu
Academiei 3/3
2028 Chișinău, Republica Moldova
Phone: +373 280-345
E-mail: tehmed@nanotech.md

Albina Nikolaeva

## Topological insulator micro wires and microlayers as potential thermoelectric materials for microelectronics

*A.Nikolaeva[1],L. Konopko[1], T. Huber[2], I. Gerishan,[1] G.Para[1]*

[1]*D.Ghitu Institute of Electronic Engineering and Nanotechnology, Chisinau, Moldova*
[2]*Department of Chemistry, Howard University, DC 20059, Washington, U.S.A.*

The best thermoelectric materials are believed to be topological insulators semiconductor $Bi_2Te3$ and Bi1-xSbx at 300K and 80K corresponding. To improve the figure of merit

$ZT=S^2\sigma T/(\chi_e +\chi_l)$ the development of new concepts - low-dimensional structures [1] and topological insulators [2] have been made. The improvement figure of merit was predicted to result either from confinement-induced increased electron density of states near the Fermi level $E_F$ or reduction in $K_l$ due to the phonon—boundary scattering or modification of the phonon spectrum.

Perfect monocrystalline of $Bi_2Te_3$ layers and Bi1-xSbx with thickness 10-20 μm were prepared using the mechanical exfoliate method by cleaving thin layer from bulk single crystal $Bi_2Te_3$ samples. X- ray diffraction studies showed that the $Bi_2Te_3$ layers were single- crystal and the plane of the layers was perpendicular to the $C_3$ trigonal axis. The micro wire in glass coating were prepared by the Ulitovsky-Taylor method.

Using p- and n- type layers as n- and p- lags in thermos element we received △T= 4˚ at 300 K on cross- section $1*10^{-4}$ $cm^2$. Applying a segmentation method (5- thermos element - increasing cross- section only to value $5*10^{-4}$ $cm^2$) we received △T=8-9˚C. It is known that an increase in the temperature of the micro- sensor by 10˚ leads to a twofold decrease the sensor durability. Our experimental samples the thermoelectric micro- coolers with efficient cooling capacity, small areas, short response time and with reproducible engineering techniques are in high demand on the telecommunication markets of the future.

This work was supported by the Moldova State Project # 20.80009.50007.02, NSF through STC CIQM 1231319, the Boeing Company and the Keck Foundation.

**Prof. Albina Nikolaeva**
Electronics of Low Dimensional Structures Lab.
D.Ghitu Institute of Electronic Engineering and Nanotechnologies
Academiei str. 3/3, MD-2028,
Chisinau, Republic of Moldova
Phone: +37322 738116
E-mail: a.nikolaeva@nanotech.md
Web: https://nanotech.md

# Multichroic Arrays of Dipole Antennas with Cold-Electron Bolometers for 210/240 GHz channels of LSPE

_A. L. Pankratov[1,2]_, E. A. Matrozova[1], D. A. Pimanov[1], A. V. Chiginev[1,2], A. V. Blagodatkin[1], A. V. Gordeeva[1,2], L. S. Kuzmin[1,3]

[1]Nizhny Novgorod State Technical University n.a. R. E. Alekseev, Nizhny Novgorod, Russia
[2]Institute for Physics of Microstructures of RAS, Nizhny Novgorod, Russia
[3]Chalmers University of Technology, Gothenburg, Sweden

The Cold-Electron Bolometer (CEB), representing SINIS structure with a hybrid superconductor/ferromagnetic (S/F) absorber, is the promising candidate for receiving systems of LSPE mission, LSPE is devoted to measuring B-modes in the cosmic microwave background in a night Arctic stratospheric flight. One of the advantages of the CEB is the effective electron cooling [1] due to the S/F absorber and normal metal traps. This allows building effective photon-noise limited receiving systems with ultimate sensitivity [2]. Due to its small size, the CEB can be effectively used to create multichroic elements for actual tasks in submillimeter astronomy due to the benefit from its ability to use co-located data [3].

According to the requirements, a receiving system of the LSPE telescope has a main frequency channel at 145 GHz for CMB measurements and two auxiliary channels of 210 and 240 GHz for cosmic dust measurements. For the moment, the basic concept of 210/240 channels is based on separate horns with TES for each frequency. Replacement of this system by a multichroic system with CEBs receiving both frequencies on-chip would improve the accuracy of co-located difference measurements of the cosmic dust.

For voltage-biased operation with a SQUID readout, the parallel arrays of about 200 dipole antennas on a 280 um Si substrate were selected. Each antenna contains a CEB as a sensitive sensor. Each array is selective to its own frequency, allowing combining the two frequency systems in one pixel. The waveguide port is located on the front side. The backside of the substrate has a thin gold layer as a backshort.

Simulation results showed good band separation and response at 210 and 240 GHz frequencies with the bandwidths of 13 and 19 GHz, and efficiency of 33 and 53%, respectively. The 145 GHz channel has a bandwidth of around 40 GHz, which is close to the requirements. The major challenge was to find the proper parameter range, where the NEP of the receiver can approach the photon NEP of the receiving signal with 6 pW power and 10 $pA/Hz^{1/2}$ SQUID noise. Detail investigations showed the impossibility of finding proper parameters for a standard CEB with two SIN tunnel junctions giving $NEP_{CEB}$ five times more than $NEP_{phot}$. These requirements can be met only with the CEB with one SIN junction and an Andreev contact (SINS structure) [4] instead of the SINIS structure. Replacement of two junctions connected in series by one junction increases responsivity by a factor of two. Replacement of two series capacitances by one decreases a volume of an absorber by a factor of four, leading to the value of $V=0.008$ $\mu m^3$. For tunnel junctions with these parameters, the resistance of R=300 Ohm and the critical temperature of 1.5 K, it is possible to get $NEP_{CEB}$ close to $NEP_{phot}=5*10^{-17}$ $W/Hz^{1/2}$.

This work was supported by the Russian Science Foundation (Project No. 21-79-20227).

**Prof. Dr. Andrey L. Pankratov**
Leading researcher,
THz Spectroscopy Dpt.,
Institute for Physics of Microstructures
Russian Academy of Sciences,
Laboratory of Superconducting Nanoelectronics
of Nizhny Novgorod State Technical University
Phone: +7 9051913223
E-mail: alp@ipmras.ru
Web: http://ipmras.ru/en/structure/people/alp

# Ultranodal pair state with Bogoliubov Fermi surfaces in FeSe$_{1-x}$S$_x$

## *T. Shibauchi*[1]
*[1]Department of Advanced Materials Science, University of Tokyo, Kashiwa, Japan*

The FeSe$_{1-x}$S$_x$ superconductors involving non-magnetic nematic phase and its quantum criticality provide a unique platform to investigate the relationship between nematicity and superconductivity [1]. It has been shown that across the nematic quantum critical point, the superconducting properties change drastically [2,3], and the non-nematic tetragonal FeSe$_{1-x}$S$_x$ ($x$>0.17) exhibits substantial low-energy states despite the high-quality of crystals. Here we have performed the muon spin rotation (μSR) measurements on FeSe$_{1-x}$S$_x$ ($x$=0, 0.20, 0.22) and observed the spontaneous internal field below the superconducting transition temperature $T_c$, providing strong evidence for time-reversal breaking (TRSB) state in bulk FeSe$_{1-x}$S$_x$ [4]. We also find that the superfluid density in the tetragonal crystals is suppressed from the expected value, indicating the presence of non-superconducting carriers. These results in FeSe$_{1-x}$S$_x$ are consistent with the recently proposed topological phase transition to a novel ultranodal pair state with Bogoliubov Fermi surface [5].

**Prof. Takasada Shibauchi**
5-1-5 Kashiwanoha,
Kashiwa, Chiba 277-8561, Japan
Phone: +81 4-7136-3774
Fax: +81 4-7136-3774
E-mail: sihbauchi@k.u-tokyo.ac.jp
Web: http://qpm.k.u-tokyo.ac.jp/shibauchi/index_e.html

## Synthesis of nZVI /PVP nanoparticles for bio – applications

*A.S. Sidorenko\*, T. D. Gutsul, E. G. Coscodan*

*Institute of Electronic Engineering and Nanotechnologies D.Ghitu ,3/3 Academiei, MD-2028 Chisinau, Republic of Moldova*

Zero-valent iron (ZVI) nanoparticles are widely studied for their exceptional properties - reduction reactions, which are mainly used in environmental protection to purify water from toxic organic solvents. [1] Earlier, we tested colloidal solutions of zero-valent iron nanoparticles stabilized with polyvinylpyrrolidone for various types of soil microorganisms. The results show that $Fe^0$/PVP nanoparticles can act as both stimulants and inhibitors of mycelial growth. The stimulating effect of $Fe^0$/PVP nanoparticles was observed in three out of five micromycete strains, namely, 1LD, 5D, and 8D. The decrease in the inhibitory effect of trifluralin on soil microorganisms Alternaria sp. 4D and P.viride is attributed to the effect of reductive degradation of Fe0/PVP nanoparticles [2] The impact of colloidal solution of zero-valent iron nanoparticles was also carried out on pathogenic microorganisms, where both an antibacterial effect and even the formation of a biofilm were found for some species. [3] In the presented work, we synthesized Zero-valent iron (ZVI) nanoparticles according to the previously described method, where hydrophilic polymers were used as stabilizers: polyethylene glycol and polyvinylpyrrolidone. The resulting nanoparticles have been characterized by X-ray powder diffraction (XRD) analysis, scanning electron microscopy (SEM), transmission electron microscopy (TEM), and FTIR spectroscopy. Alternaria alternate was chosen as the object of biological study, Birolatis sorokiana is a harmful helminthosporious type of „black germ" of wheat. In case of helminthosporic infection, fungal mycelium penetrates into the embryo and shit of wheat seeds, which causes the formation of shriveled seeds with low germination, and during germination root rot. Experiments on germination of contaminated seeds and treatment with colloidal solutions of nanoparticles were carried out in a growing chamber at a temperature of 23C, control samples were contaminated seeds in distilled water in accordance with the standards GOST 12038-84, ISO 22030-2009. The concentration of colloidal solutions was in the range of 0.5 mg / L - 100 mg / l. As a result of the experiment, laboratory seed germination increased by 27.3% in comparison with the control sample.

**Prof. Dr. Anatolie Sidorenko**
Institute of Electronic Engineering and Nanotechnologies "D. GHITU"
Academiei 3/3, Kishinev MD2028 Moldova
Tel +37322-727072; FAX +37322-727088
E-mail: anatoli.sidorenko@kit.edu
Web: http://nanotech.md
ORCID 0000-0001-7433-4140
Linked In: https://www.linkedin.com/in/anatoli-sidorenko-102a1629/

## Study of a new generation of rockets for active influence on clouds

*E.A. Zasavitsky, D. I. Karagenov and A. S. Sidorenko*

*D.Ghitu Institute of Electronic Engineering and Nanotechnologies, Chisinau, Republic of Moldova*

The technology is based on the use of a small aerodynamic stand, which makes it possible to test the yield of various rockets for active impacts on clouds, in particular, rockets with a propulsion engine that operates throughout the entire flight path and uses a new type of solid propellant. These rockets can significantly increase the yield of active crystallization centers per unit length of the seeding path.

An experimental verification of the yield of active crystallization centers per gram of the composition of rockets on solid fuel generating ice-forming nuclei has been conducted at Ghitu IIEN on an upgraded stand, which makes it possible to test the rockets under conditions that closely simulate the flight conditions. The verification has confirmed the advantages of the rockets and shown that the yield of active ice-forming nuclei during the combustion of full-sized mid-flight rocket engines is ~$10^{14}$ g$^{-1}$ at a supercooled model fog temperature of -$10^{0}$C.

The tests conducted on a stand for testing rockets on solid fuel generating ice-forming nuclei have shown that, compared with a conventional anti-hail rocket, the tested rockets can significantly increase the yield of active crystallization centers. It has been shown that use of rockets on solid fuel generating ice-forming nuclei provides the high-efficiency seeding of hail-hazardous clouds with artificial nuclei and, as a consequence, the suppression of hail-formation processes in potentially hazardous clouds. It is significant that the aerosol is characterized not only by a high particle yield, but also an extremely high temperature threshold for crystallization (about -$4^{0}$C). This fact suggests that a fairly high yield of active crystals in the above temperature region will make it possible to implement active impacts to artificially increase precipitation and dissipate clouds.

**Dr. Zasavitsky E. A.**
Academiei str. 3/3, Chisinau,
MD-2028 Republic of Moldova
E-mail: efim@nanotech.md
Phone: +373-22-73-71-97
Fax: +373-22-72-70-88
Web: http://www.nanotech.md

## Environmental aspects of long-term hail-suppression activities in Moldova

*E. A. Zasavitsky, E. I. Potapov and A. S. Sidorenko*
*D.Ghitu Institute of Electronic Engineering and Nanotechnologies, Chisinau, Republic of Moldova*

The results of studies of the lead and silver content in water bodies and air and the ice nuclei concentrations in the regions subjected to hail protection in the Republic of Moldova in 1977÷1991 are summarized. Until 1983, hail-hazardous clouds were seeded mostly with crystallizing pyrotechnic compositions based on lead iodide; subsequently, silver iodide was used. The problem of environmental pollution in the regions involved in the activities using various crystallizing reagents based on heavy metals are discussed. Accumulation of these metals depending on the type and lifetime of the reagents used is estimated.

Long-term observations of the use of AgI-based ice-forming compositions have not revealed either a tendency of silver aerosol accumulation in the ground air or its correlation with the amount of the consumed reagent. The analysis of results clearly show that there are no environmentally harmful effects arising from hail-suppression cloud seeding program involving silver iodide aerosols in Moldova.

**Dr. Zasavitsky E. A.**
D.Ghitu Institute of Electronic Engineering and Nanotechnologies
Academiei str. 3/3, Chisinau,
MD-2028 Republic of Moldova
E-mail: efim@nanotech.md efimzasavitsky@gmail.com
Phone: +373-22-73-71-97
Fax: +373-22-72-70-88
Web: http://www.nanotech.md

# AUTHOR INDEX

| | |
|---|---|
| Fiodor M. Muntyanu | 58, 64 |
| Iurie N. Nica | 76 |
| Albina Nikolaeva | 74, 78 |
| Klenov Nikolai | 41, 43 |
| Andrey L. Pankratov | 26, 33, 53, 79 |
| Dmitry A. Pimanov | 53, 79 |
| Ilhom Rahmonov | 37, 59 |
| Leonid S. Revin | 26, 33, 49 |
| Valeriy Ryazanov | 40 |
| Yuri B. Savva | 25, 43 |
| Olesya Yu. Severyukhina | 60 |
| Takasada Shibauchi | 28, 81 |
| Yury M. Shukrinov | 37, 52, 59 |
| Anatolie Sidorenko | 43, 46, 58, 60, 67, 68, 71, 76, 82, 83, 84 |
| Andrei Sirbu | 65, 68 |
| Stepan V. Suvorov | 67 |
| Alexander V. Vakhrushev | 36, 43, 60, 67 |
| Zihan Wei | 11 |
| Silke Wolter | 55 |
| Igor V. Yanilkin | 42, 61, 62 |
| Roman V. Yusupov | 42, 61, 62 |
| E. A. Zasavitsky | 83, 84 |